\documentclass[utf8]{pepe} 

\setcitestyle{square} 
\usepackage{url,hyperref,lineno,microtype,subcaption}
\usepackage[onehalfspacing]{setspace}

\newcommand{\fet}[1]{\mbox{\boldmath $#1$}}
\newcommand{\beq}{\begin{equation}}
\newcommand{\eeq}{\end{equation}}
\newcommand{\beqa}{\begin{eqnarray}}
\newcommand{\eeqa}{\end{eqnarray}}
\newcommand{\nn}{\nonumber \\ }

\def\keyFont{\fontsize{8}{11}\helveticabold }
\def\firstAuthorLast{E.~Epelbaum {et~al.}} 

\def\Authors{E.~Epelbaum\,$^{1}$, H.~Krebs\,$^{1}$ and P.~Reinert\,$^{1}$}

\def\Address{$^{1}$  Ruhr-Universit\"at Bochum, Fakult\"at f\"ur Physik und
        Astronomie, Institut f\"ur Theoretische Physik II,  D-44780
        Bochum, Germany}

\def\corrAuthor{}
\def\corrEmail{$\,$ evgeny.epelbaum@rub.de, hermann.krebs@rub.de, patrick.reinert@rub.de}

\begin{document}
\onecolumn
\firstpage{1}

\title[High-precision nuclear forces from chiral EFT]{High-precision nuclear forces from chiral EFT:
State-of-the-art, challenges and outlook}

\author[\firstAuthorLast ]{\Authors} 
\address{} 
\correspondance{} 

\extraAuth{}

\maketitle

\begin{abstract}

\section{}
We review a new generation of nuclear forces derived  in chiral
effective field theory using the recently proposed  semilocal
regularization method.  We outline the conceptual foundations
of nuclear chiral effective field theory, discuss all steps needed to compute nuclear observables
starting from the effective chiral Lagrangian and consider selected
applications in the two- and few-nucleon sectors.
We highlight key challenges in
developing high-precision tree-body forces, such as the need to 
maintain consistency between two- and many-body interactions 
and constraints placed by the chiral and gauge symmetries after
regularization. 

\tiny
 \keyFont{ \section{Keywords:} Nuclear forces, effective field theory, chiral perturbation theory, regularization, few-body systems} 
\end{abstract}

\section{Introduction}

Almost thirty years ago, Weinberg put forward his groundbreaking idea
to apply chiral
perturbation theory (ChPT), the low-energy effective field theory (EFT) of
QCD, to the derivation of nuclear interactions \cite{Weinberg:1990rz,Weinberg:1991um}. This seminal
work has revolutionized the whole field of nuclear physics by
providing a solid theoretical basis and offering a
systematically improvable approach to low-energy nuclear
structure and reactions.  

So where do we stand today in the implementation of the program
initiated by Weinberg? Much has been learned about specific features
of the nuclear interactions and currents and about the role of many-body forces
from the point of view of the effective chiral Lagrangian, see Refs.~\cite{Epelbaum:2005pn,Epelbaum:2008ga,Machleidt:2011zz,Hammer:2019poc}
for review articles covering different research areas,
while some issues are still under debate \cite{Epelbaum:2018zli,Hammer:2019poc}.  Meanwhile,
the interactions derived in chiral EFT, sometimes referred to as ``chiral
forces'', have largely replaced phenomenological potentials developed
in the nineties of the last century. They are nowadays 
commonly used in \emph{ab initio} nuclear structure calculations, see
Refs.~\cite{Epelbaum:2018ogq,Piarulli:2017dwd,Lonardoni:2017hgs,Hagen:2016uwj,Gebrerufael:2016xih,Cipollone:2014hfa} 
for recent examples using a variety of continuum \emph{ab initio}
methods and \cite{Epelbaum:2011md,Elhatisari:2015iga,Lahde:2019npb} for selected
highlights from nuclear lattice simulations. With the most recent
chiral nucleon-nucleon (NN) 
potentials \cite{Reinert:2017usi} providing a nearly perfect description of the
mutually consistent neutron-proton (np) and proton-proton (pp) scattering data
below pion production threshold from the
Granada-2013 database \cite{Perez:2013jpa}, the
two-nucleon sector is already in a very good
shape. On the other hand, three-nucleon forces (3NF) are much less understood at the
quantitative level \cite{KalantarNayestanaki:2011wz} and constitute an important frontier in nuclear
physics \cite{Hammer:2012id}. 

In this article we focus on the latest generation of chiral nuclear
forces based on an improved regularization approach \cite{Epelbaum:2014efa,Epelbaum:2014sza,Reinert:2017usi}, which allows one
to maintain the long-range part of the interaction. We review our
recent work along these lines in the two-nucleon sector, describe
the ongoing efforts by the Low-Energy Nuclear Physics International
Collaboration (LENPIC) towards developing consistent many-body forces
and solving the structure and reactions of light nuclei, and discuss
selected applications.

Our paper is organized as follows. In section
\ref{sec2}, we outline the foundations of the employed theoretical
framework.  Section~\ref{sec3} gives an overview of various methods
to derive nuclear forces and currents from the effective
chiral Lagrangian. It also summarizes the available results for
nuclear potentials derived 
using dimensional regularization (DR). In section~\ref{sec4} we
present the improved semilocal regularization approach, which is
utilized in the most accurate and precise NN potentials of Ref.~\cite{Reinert:2017usi}. We also
discuss the challenges that need to be addresses to construct \emph{consistently regularized} 3NFs
and exchange current operators beyond tree level, which are 
not restricted to any particular type of cutoff regularization. Section~\ref{sec5} is
devoted to uncertainty quantification in chiral EFT. Selected results
for the NN system, three-nucleon scattering and light nuclei are
presented in section~\ref{sec6}. We conclude 
with a short summary and outlook in section~\ref{sec7}.

\section{The framework in a nutshell}
\label{sec2}

Throughout this work, we restrict ourselves to the two-flavor case of
the light up- and down-quarks and employ the simplest version of the effective
chiral Lagrangian with pions and nucleons as the only
active degrees of freedom. Contributions of the $\Delta$(1232) isobar
to the nuclear potentials are discussed in
Refs.~\cite{Ordonez:1995rz,Kaiser:1998wa,Krebs:2007rh,Epelbaum:2007sq,Epelbaum:2008td,Krebs:2018jkc}.
The effective
Lagrangian involves all possible interactions between pions and
nucleons compatible with the symmetries of QCD and is organized in
powers of derivatives and quark (or equivalently pion) masses. Pions
correspond to the (pseudo) Nambu-Goldstone bosons of the spontaneously broken
axial generators and thus transform nonlinearly with respect to 
chiral SU(2)$_L$$\times$SU(2)$_R$ transformations. The effective
Lagrangian can be constructed in a straightforward way using covariantly transforming
building blocks defined in terms of the pion fields
\cite{Coleman:1969sm,Callan:1969sn}. All applications reviewed in this
paper rely on a nonrelativistic treatment of the nucleon fields and
make use of the heavy-baryon formalism to eliminate the 
nucleon mass $m$ from the leading-order Lagrangian.
The individual terms in the
effective Lagrangian are multiplied by the corresponding coupling
constants, commonly referred to as low-energy 
constants (LECs), which are not fixed by the symmetry and typically need to
be determined from experimental data. The most accurate currently
available nuclear potentials at fifth order in the chiral expansion,
i.e.~at N$^4$LO,
require input from the following effective Lagrangians (with each line
containing the contributions with a fixed number of the nucleon fields) 
\begin{eqnarray}
  \label{Lagr}
\mathcal{L}_{\rm eff} &=& \mathcal{L}_{\pi}^{(2)} (M_\pi , F_\pi ) +
                          \mathcal{L}_{\pi}^{(4)} (l_{1, \ldots ,
                          7})\nn
                          &+& \mathcal{L}_{\pi \rm  N}^{(1)} (g_A ) +
                              \mathcal{L}_{\pi \rm N}^{(2)} (m, c_{1,
                              \ldots , 7} ) +
                              \mathcal{L}_{\pi \rm N}^{(3)} (d_{1,
                              \ldots , 23} ) +   \mathcal{L}_{\pi \rm N}^{(4)} (e_{1,
                              \ldots , 118} ) \nn
                              &+& \mathcal{L}_{\rm N N}^{(0)} (C_S, C_T) +
                                  \mathcal{L}_{\rm N N}^{(2)} (C_{1,
                                  \ldots , 7})  +
                                  \mathcal{L}_{\rm N N}^{(4)} (D_{1,
                                  \ldots , 12})  + \mathcal{L}_{\pi
                                  \rm 
                                  NN}^{(1)} (D )  + \ldots \nn
                                  &+&
                                  \mathcal{L}_{\rm N N N}^{(0)} (E ) +
                                      \mathcal{L}_{\rm N N N}^{(2)} (E_{1,
                                      \ldots , 10} )\,,
  \end{eqnarray}
  where $M_\pi$ and $F_\pi$ are the pion mass and decay
  constant\footnote{Strictly speaking, $M_\pi$ is to be understood as
    the pion mass to leading order in the chiral expansion while
    $F_\pi$ and other parameters in the effective Lagrangian refer to
    the corresponding LECs in the chiral limit of vanishing light
    quark masses.},
  $g_A$ is the nucleon axial-vector coupling while
$l_i$, $c_i$,  $d_i$, $e_i$, $C_i$, $D$, $D_i$,  $E$ and $E_i$ are
further LECs. The superscript $n$ of $\mathcal{L}^{(n)}$ denotes the
number of derivatives and/or $M_\pi$-insertions and is sometimes
referred to as the chiral dimension. Notice that we only show new LECs that appear in the
corresponding Lagrangians and suppress the dependence on the LECs
appearing at lower orders. The pionic Lagrangian can be found
in Ref.~\cite{Gasser:1983yg},  $\mathcal{L}_{\pi \rm
  N}$ is given in  Refs.~\cite{Bernard:1995dp,Fettes:2000gb}, 
$\mathcal{L}_{\rm N N}^{(0)}$ was introduced in
Refs.~\cite{Weinberg:1990rz,Weinberg:1991um},  $\mathcal{L}_{\rm N N}^{(2)}$
can be found in Refs.~\cite{Epelbaum:2001fm,Girlanda:2010ya},
the minimal form of $\mathcal{L}_{\rm N N}^{(4)}$ is given in
Ref.~\cite{Reinert:2017usi}, $\mathcal{L}_{\pi \rm N N}^{(1)}$ and
$\mathcal{L}_{\rm N N N}^{(0)}$ are discussed in
Ref.~\cite{Epelbaum:2002vt} while $\mathcal{L}_{\rm N N N}^{(2)}$ was
constructed in Ref.~\cite{Girlanda:2011fh}. Notice further that the chiral
symmetry breaking terms $\propto M_\pi^2$ are not shown explicitly 
in $\mathcal{L}_{\rm N N}$ and $\mathcal{L}_{\rm N N N}$. For
calculations at the physical value of the quark masses, their
contributions are absorbed into the LECs listed in
Eq.~(\ref{Lagr}). We have, furthermore, restricted ourselves in this equation to
isospin-invariant terms for the Lagrangians involving two and three
nucleons. The single-nucleon Lagrangian  $\mathcal{L}_{\pi \rm N}$
does involve isospin-breaking contributions due to the quark mass
difference and can be extended to include virtual photon effects
Refs.~\cite{Muller:1999ww,Gasser:2002am}. The ellipses in
the second-to-last line of Eq.~(\ref{Lagr}) refer to higher-order
Lagrangians $\mathcal{L}_{\pi \rm NN}$, which have not been worked out
yet and would be needed to finalize the derivation of the
3NF at N$^4$LO. 

The long-range parts of the nuclear forces emerge from pion exchange
diagrams and can be derived from $\mathcal{L}_{\pi}$ and
$\mathcal{L}_{\pi \rm N}$. Fortunately, only a very restricted set of
(linear combinations of) LECs from these Lagrangians contributes to the
$\pi N \to \pi N$ and $\pi N \to \pi \pi N$ scattering amplitudes, which
enter as subprocesses when deriving the long-range nuclear
interactions up to N$^4$LO, namely $c_{1 , \ldots , 4}$ from
  $\mathcal{L}_{\pi \rm N}^{(2)}$, $d_1 + d_2$,  $d_{3, 5, 18}$ and
  $d_{14} - d_{15}$ from $\mathcal{L}_{\pi \rm N}^{(3)}$ and
  $e_{14,\ldots ,18}$ from $\mathcal{L}_{\pi \rm N}^{(4)}$. Here, we made use of the fact that the
  contributions from the LECs $l_3$, $e_{19, \ldots , 22}$ and $e_{35,
    \ldots , 38}$ can be absorbed into the appropriate shifts of the
  LECs $c_i$  \cite{Krebs:2012yv}. All these $\pi$N LECs can nowadays be
  reliably extracted by matching the $\pi$N scattering amplitude from
  the recent Roy-Steiner equation analysis \cite{Hoferichter:2015hva} with 
  ChPT at the subthreshold point \cite{Hoferichter:2015tha}, see also
  Ref.~\cite{Siemens:2017opr} for an alternative strategy. Thus, \emph{the
  long-range nuclear interactions are completely determined by the
  spontaneously broken approximate chiral symmetry of QCD and
  experimental/empirical information on the $\pi$N system in a
  parameter-free way}. The two- and three-nucleon
interactions in the last two lines of Eq.~(\ref{Lagr}) parametrize the
short-range part of the nuclear forces, and the corresponding LECs
have to be determined from NN scattering and three- or
more-nucleon observables. 

In the single-nucleon sector, the effective Lagrangian
$\mathcal{L}_\pi + \mathcal{L}_{\pi \rm N}$ can be used to
systematically compute the scattering amplitude in perturbation theory
by applying the chiral expansion, a simultaneous expansion in
particles' external three-momenta $p \equiv | \vec p \, |$ and around the chiral limit 
$M_\pi \to 0$. The importance of every Feynman diagram is estimated
by counting powers of the soft scales
and applying the rules of naive dimensional analysis (NDA). 
The expansion parameter  $Q \in \{
p/\Lambda_{\rm b}, \, M_\pi / \Lambda_{\rm b}\}$ is determined by the breakdown
scale $\Lambda_{\rm b}$, which may (optimistically) be expected to be of the order of the
$\rho$-meson mass.\footnote{An upper bound for $\Lambda_{\rm b}$ is
  set by the scale $4 \pi F_\pi$ emerging from pion loops \cite{Manohar:1983md}.} At every order in
the chiral expansion only a finite number of Feynman diagrams need to be
evaluated. For more details on ChPT in the
1N sector see the review article
\cite{Bernard:2007zu}. 

Contrary to the 1N case, the NN S-wave scattering amplitude
exhibits poles in the near-threshold region corresponding to the
bound state (deuteron) and the virtual state in the $^1$S$_0$
channel, which signal the breakdown of perturbation theory. In this
context, it was pointed out by Weinberg that the contributions of
multi-nucleon ladder diagrams are enhanced compared to the estimation
based on the chiral power counting due to the appearance of pinch
singularities (in the $m \to \infty$ limit) \cite{Weinberg:1990rz,Weinberg:1991um}. Weinberg also argued
that the nucleon mass needs to be counted as $m \sim \Lambda_{\rm b}^2/M_\pi
\gg \Lambda_{\rm b}$  in order to formally justify the need to perform a
nonperturbative resummation of the ladder contributions. Given that the
ladder diagrams are automatically resummed by solving the few-nucleon
Schr\"odinger equation,
Weinberg's chiral EFT approach to low-energy nuclear systems, perhaps
not surprisingly, resembles the quantum mechanical $A$-body problem
\beq
\label{Schroed}
\bigg[ \bigg( \sum_{i=1}^A \frac{- \Delta_i}{2 m} + \mathcal{O}
\big( m^{-3} \big) \bigg) + V_{\rm 2N} + V_{\rm 3N} + V_{\rm 4N} + \ldots \bigg]
| \Psi \rangle = E | \Psi \rangle\,,
\eeq
where $\Delta_i$ is the Laplace operator acting on the nucleon $i$. 
The nuclear potentials $V_{\rm 2N}$, $V_{\rm 2N}$, $\ldots$  receive
contributions from diagrams that cannot be reduced to ladder
iterations and are calculable in a systematically improvable way
within ChPT.

\begin{figure}[tb]
\begin{center}
\includegraphics[width=0.65\textwidth]{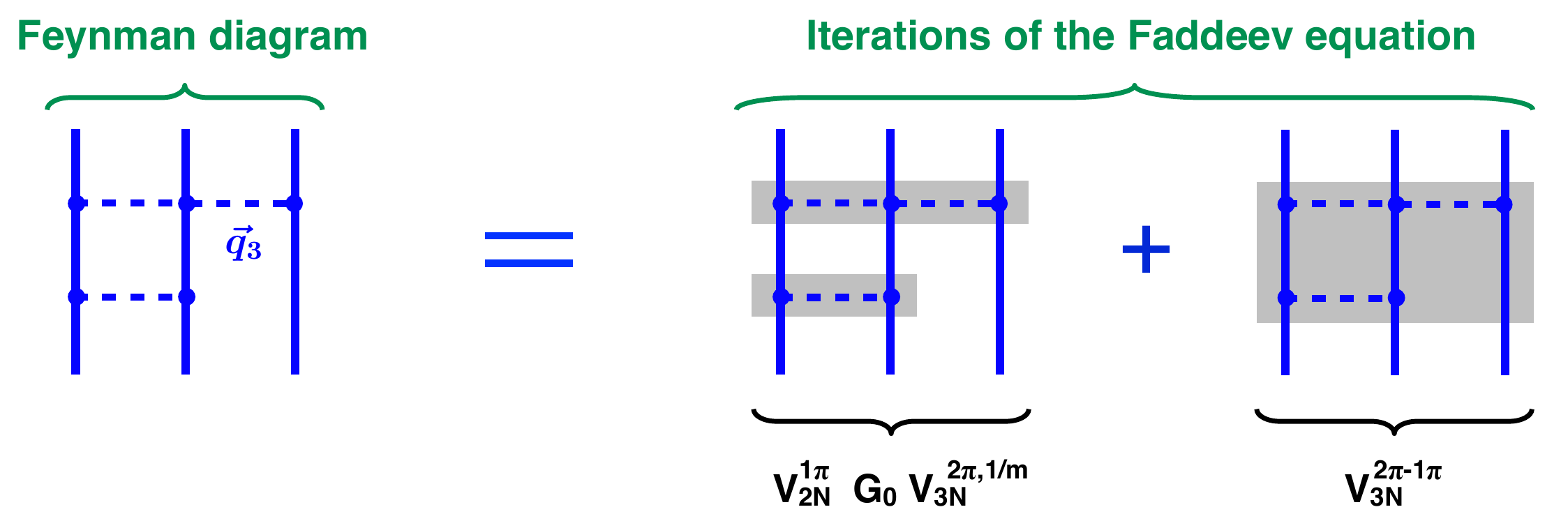}
\end{center}
\caption{Representation of the on-shell scattering amplitude from the one-pion-two-pion-exchange
  Feynman diagram (left)  in terms of iterations of the Faddeev
  equation (right). Gray-shaded rectangles visualize the corresponding
  two- and three-nucleon potentials $V_{\rm 2N}^{1 \pi}$,  $V_{\rm
    3N}^{2 \pi, \, 1/m}$ and   $V_{\rm
    3N}^{2 \pi - 1 \pi}$ while $G_0$ denotes the free resolvent operator for 
  nonrelativistic nucleons.}\label{fig:1}
\end{figure}

Among the many attractive features, the approach outlined above allows one
to maintain \emph{consistency} between nuclear forces and exchange
current operators which are scheme-dependent quantities. To
illustrate the meaning and importance of consistency consider the
Feynman diagram on the left-hand side (l.h.s.) of the equality shown in
Fig.~\ref{fig:1} as an example. The corresponding
(on-shell) contribution to the scattering amplitude features both a
reducible (i.e.~of a ladder-type) and irreducible pieces as visualized in
the figure. Reducible contributions to the amplitude are
resummed up to an infinite order when solving the Faddeev equation
corresponding to Eq.~(\ref{Schroed}). In doing so, the  diagrams
corresponding to its zeroth and first
iterations shown in the figure \emph{must} match
the result obtained from the Feynman diagram when taken on the energy shell. The
iterative contribution from the first graph on the right-hand side of
the depicted equality, however, involves NN and 3N potentials, whose off-shell
behavior is scheme dependent. Also the 3NF corresponding to the last
diagram is scheme dependent (even on the energy shell) \cite{Bernard:2007sp}, and only
a \emph{consistent} choice of the involved two- and three-nucleon
potentials guarantees the validity of matching for 
the scattering amplitude. This can indeed be verified explicitly
using the expressions for the 
3NFs $V_{\rm  3N}^{2 \pi, \, 1/m}$ from Eqs.~(4.9)-(4.11) 
of Ref.~\cite{Bernard:2011zr} and $V_{\rm  3N}^{2 \pi - 1 \pi}$ from Eqs.~(2.16)-(2.20) of 
Ref.~\cite{Bernard:2007sp} and employing DR to
evaluate loop integrals.\footnote{Notice that the 
  contributions from diagrams shown in Fig.~\ref{fig:1} are finite in DR.} 

Clearly, DR is impractical for a numerical
solution of the  $A$-body problem and is usually
replaced by cutoff regularization. Renormalization of the
Schr\"odinger equation in the context of chiral EFT is a controversial
and heavily debated topic, see
Refs.~\cite{Lepage:1997cs,PavonValderrama:2004nb,Nogga:2005hy,Birse:2005um,Epelbaum:2006pt,Epelbaum:2009sd,Long:2011xw,Valderrama:2016koj,Epelbaum:2018zli,Hammer:2019poc}
for a range of 
opinions. The essence of the problem is related
to the nonrenormalizable nature of the Lippmann-Schwinger (LS) equation for
NN potentials truncated at a finite order in the chiral
expansion. Except for the leading-order (LO)  equation in pionless EFT
and in chiral  EFT in spin-singlet channels, ultraviolet (UV) divergences
emerging from the loop expansion of the scattering amplitude cannot be
absorbed into redefinitions of parameters appearing in the
truncated potentials \cite{Epelbaum:2018zli,Savage:1998vh}. The problem can be avoided by treating the
one-pion exchange (OPE) and higher-order contributions to the potential in 
perturbation theory using e.g.~the systematic power counting scheme proposed by Kaplan,
Savage and Wise \cite{Kaplan:1998tg}, but the resulting approach unfortunately fails to converge
(at least) in certain spin-triplet channels \cite{Cohen:1999iaa,Fleming:1999ee}, see also
Refs.~\cite{Kaplan:2019znu} for a recent discussion. A
renormalizable framework with the one-pion exchange potential (OPEP) treated
nonperturbatively was proposed in \cite{Epelbaum:2012ua} based on a manifestly
Lorentz invariant form of the effective Lagrangian. This approach
requires a perturbative inclusion of higher-order contributions to the
potential in order to maintain renormalizability
(which may lead to convergence issues in some channels
\cite{Epelbaum:2015sha}) but has not been systematically explored
beyond LO yet.   

Throughout this work we employ a finite-cutoff version of nuclear
chiral EFT in the formulation of Ref.~\cite{Lepage:1997cs}, which is
 utilized in most of the applications available today. This is so
far the only scheme, that has been advanced to high
chiral orders and successfully applied to a broad range of few- and
many-nucleon systems. Below, we briefly summarize the basic
steps involved in the calculation of nuclear observables within this
framework. In the following sections, all four steps outlined below will be
discussed in detail. 
\begin{itemize}
\item[i.] {\it Derivation of nuclear forces and current operators from
  the effective chiral Lagrangian.}  This can be achieved by separating
out irreducible contributions to the $A$-nucleon scattering amplitude that cannot
be generated by iterations of the dynamical equation using various
methods outlined in section~\ref{sec3}. The derivations are
carried out in perturbation theory using the standard chiral power
counting. In contrast to ChPT for the scattering
amplitude, special efforts are needed to arrive at renormalized nuclear
potentials. This requires that all UV divergences from
\emph{irreducible} loop diagrams are cancelled by the corresponding counter terms.
The renormalizability requirement imposes strong constraints on the unitary
ambiguity of nuclear forces and currents \cite{Krebs:2012yv,Epelbaum:2006eu,Epelbaum:2007us,Kolling:2011mt,Krebs:2016rqz}.  
\item[ii.] {\it Introduction of a regulator for external (off-shell)
    momenta of the nucleons} in order to make the $A$-body Schr\"odinger
  equation well-behaved. Given the lack of counter terms needed to
  absorb all UV divergences from iterations of the dynamical equation
  with a truncated potential, the
  (momentum-space) cutoff $\Lambda$ must not be set to arbitrarily high
  values but should be kept of the order of the breakdown scale,
  $\Lambda \sim \Lambda_{\rm b}$
  \cite{Lepage:1997cs,Epelbaum:2009sd,Epelbaum:2018zli}. The
  accessible cutoff window is, in practice, 
  further restricted by the need to avoid the appearance of spurious
  deeply bound states which provide a severe complication for applications
  beyond the NN system \cite{Epelbaum:2002ji} and a preference for soft interactions in
  order to optimize convergence of {\it ab initio} many-body methods.   
  Given the rather restricted available cutoff window, it is important
  to employ regulators that minimize the amount of finite-cutoff
  artifacts, see section~\ref{sec4} for discussion. While the
  regulator choice for $V_{\rm 2N}$ still features a high degree of
  ambiguity, maintaining
the relevant symmetries and  consistency with \emph{regularized} many-body forces and exchange
  currents beyond tree level represents a highly nontrivial task \cite{Krebs:2019uvm,Epelbaum:2019jbv}, see section
  \ref{sec4} for an example and discussion. 
\item[iii.] {\it Renormalization of the few-nucleon amplitude by fixing the
   short-range multi-nucleon interactions from low-energy
   experimental data}, see section~\ref{sec6} for details.  This
 allows one to express the calculated scattering amplitude in
  terms of observable quantities instead of the  \emph{bare} LECs
    $C_{S,T} (\Lambda )$,  $C_{i} (\Lambda )$, $D_{i} (\Lambda )$, $D
    (\Lambda )$, $E (\Lambda )$, $E_i (\Lambda )$, $\ldots$, and
    amounts to implicit renormalization of the amplitude. Notice
    that in the pion and 1N sectors of ChPT,
    renormalization is usually carried out \emph{explicitly} by splitting the bare LECs
    $l_i$, $d_i$, $e_i$, $\ldots$, into the (finite) renormalized ones and
    counter terms, e.g.~$d_i = d_i^{\rm r} (\mu ) + R_i (\mu)$. 
    Here, $\mu$ denotes the renormalization scale while $R_i$ are the
    corresponding counter terms, which diverge in the limit of a
    removed regulator (i.e.~$\Lambda \to \infty$ in the cutoff
    regularization or the number of dimensions $d \to 4$ in DR).
    Such a splitting is not unique as reflected by the scale $\mu$,
    and the appropriate
    choice of renormalization conditions is essential to 
    maintain the desired power counting, i.e.~to ensure the appropriate
    scaling behavior of  
    renormalized contributions to the amplitude leading to a
    systematic and self-consistent scheme, see e.g.~Refs.~\cite{Gegelia:1999qt,Fuchs:2003qc}. In the
    few-nucleon sector, the nonperturbative
    resummation of pion-exchange potentials via Eq.~(\ref{Schroed})
    can only be carried out numerically\footnote{See, however,
      Ref.~\cite{Kaplan:2019znu} for analytical results in the chiral limit.}, which
    leaves the \emph{implicit} renormalization outlined above as the
    only available option. Notice that contrary to the renormalized LECs $l_i^{\rm r} (\mu )$,
    $d_i^{\rm r} (\mu)$,   $\ldots$, the bare LECs $C_{S,T} (\Lambda
    )$,  $C_{i} (\Lambda )$, $\ldots$, must be re-determined at every
    order in the expansion. 
\item[iv.] {\it Estimation of the truncation uncertainty and
  a-posteriori  consistency checks of the obtained results.}  These include, among
  others, testing the naturalness of the extracted LECs \cite{Reinert:2017usi}, making error plots for phase shifts as
  suggested in Refs.~\cite{Lepage:1997cs,Griesshammer:2015osb}, verifying a reduced residual
  $\Lambda$-dependence of observables (within a specified cutoff
  range) upon including higher-order short-range interactions, see e.g.~Fig.~4 of Ref.~\cite{Epelbaum:2015pfa}, and confronting the
  contributions of many-body interactions and/or exchange currents with
  estimations based on the assumed power counting \cite{Binder:2015mbz,Epelbaum:2019zqc}. Our approach to error analysis
  is outlined in section~\ref{sec5}, while selected consistency
  checks are discussed in section~\ref{sec6}. 
\end{itemize} 

Before closing this section, several remarks are in order. First, 
we emphasize that the approach outlined above is applicable at
the physical quark masses. Quark mass dependence of nuclear
observables can be studied more efficiently 
in the renormalizable chiral EFT framework of
Refs.~\cite{Epelbaum:2012ua,Epelbaum:2013ij}, see also 
Refs.~\cite{Baru:2015ira,Baru:2016evv,Lahde:2019yvr} for an alternative
method. Secondly, the validity (in the EFT sense) of the finite-cutoff
EFT formulation outlined above has been demonstrated numerically
by means of the
error plots \cite{Lepage:1997cs,Epelbaum:2010nr}  and
analytically \cite{Epelbaum:2009sd} for toy-models with long-range interactions. 
It can also be easily verified in pionless EFT. For the case of exactly known
non-singular long-range potentials, the employed approach reduces in the NN
sector to  the well-known
modified effective range expansion \cite{vanHaeringen:1981pb}. 
The relation between
the choice of renormalization 
conditions and power counting is discussed within pionless EFT in
Ref.~\cite{Epelbaum:2017byx}.\footnote{For pionless EFT or chiral EFT
  with perturbative pions, the NN amplitude can be
  calculated analytically, and renormalization can be carried out
  explicitly.}  That paper provides an explicit example of the choice
of  subtraction scheme (i.e.~renormalization conditions), which leads to a self-consistent EFT
approach for two particles with both a natural and unnaturally large
scattering length, while respecting the NDA scaling of renormalized
LECs.
Notice that in all applications reviewed in this article, few-nucleon
short-range interactions are counted according to NDA. A number of
authors advocate alternative approaches, in particular by inferring the importance of
short-range operators from the requirement of $\Lambda$-independence
of the scattering amplitude at arbitrarily large values of $\Lambda$ as articulated in
detail in Ref.~\cite{Hammer:2019poc}. However, performing the loop
expansion of the solution of the LS equation in spin-triplet channels for the
resummed OPEP shows that the scattering amplitude
is only
\emph{partially} renormalized in spite of the fact that it admits, in
some cases, a finite $\Lambda \to
\infty$ limit at a fixed energy \cite{Epelbaum:2018zli}. The danger of choosing $\Lambda \gg
\Lambda_{\rm b}$ in such partially renormalized nonperturbative expressions
is demonstrated using an exactly solvable model in Ref.~\cite{Epelbaum:2009sd}.

\section{Chiral perturbation theory for nuclear potentials}
\label{sec3}

One method to decouple
pion-nucleon and purely nucleonic subspaces of the Fock space, thereby reducing
a quantum field theoretic problem to a quantum mechanical
one, is the unitary transformation (UT) technique.
Let $\eta$ and $\lambda$ be the projection operators onto the
purely nucleonic subspace of the Fock space and the rest, 
respectively. The time-independent Schr\"odinger equation can be written in the
form
\begin{eqnarray}
 \left(
\begin{array}{rr}
\eta\,H\,\eta & \eta\, H\, \lambda \\
\lambda\, H\,\eta & \lambda\,H\,\lambda\\
\end{array}
\right)
\left(
\begin{array}{r}
\eta\,|\Psi\rangle\\
\lambda\,|\Psi\rangle\\
\end{array}
\right)
&=& 
E \left(
\begin{array}{r}
\eta\,|\Psi\rangle\\
\lambda\,|\Psi\rangle
\end{array}\right),\label{schroedinger_eq}
\end{eqnarray}
where $E$ denotes the eigenenergy of the $\pi$N system.
The idea is to apply a UT to the Hamilton operator $H$
in order to block diagonalize the matrix on the l.h.s.~of
Eq.~(\ref{schroedinger_eq}) leading to 
\begin{eqnarray}
\left[U^\dagger H\,U\right]U^\dagger|\Psi\rangle &=&
                                                 E\,U^\dagger|\Psi\rangle.\label{schroedinger_eq_rotated}
\end{eqnarray}
The decoupling requirement is given by 
\begin{eqnarray}
&&\eta\,U^\dagger H\,U\,\lambda \,=\, \lambda\,U^\dagger H\,U\,\eta \,=\,0.\label{decoupling_eq}
\end{eqnarray}
To construct the UT $U$ we first introduce a M{\o}ller
operator $\Omega$~\cite{Suzuki:PTP1983}, which is defined by
\beqa
|\Psi\rangle&=&\Omega \eta |\Psi\rangle\label{moeller:op:def}
\eeqa
with the requirement 
\beqa
\Omega &=&\Omega  \eta.\label{moeller:eta:right}
\eeqa
The M{\o}ller operator reproduces the original state out of projected
state. By projecting Eq.~(\ref{moeller:op:def}) onto the model space $\eta$ one
obtains the identity
\beqa
\eta \Omega &=& \eta.\label{moeller:eta}
\eeqa
Using Eq.~(\ref{moeller:op:def}), we can write the time-independent Schr\"odinger equation in the
form 
\beq
\big(E-H_0\big) \Omega \eta |\Psi\rangle= V
|\Psi\rangle,\label{HOmegaSchroedinger}
\eeq
where $H_0$ denotes a free Hamiltonian. On the other hand, 
projecting the
original Schr\"odinger equation Eq.~(\ref{schroedinger_eq}) onto
the model space and applying on the resulting equation the operator
$\Omega$, we obtain 
\beq
\big(E \Omega-\Omega H_0\big) \eta |\Psi\rangle= \Omega \eta V
|\Psi\rangle .\label{OmegaHSchroedinger}
\eeq
Subtracting Eq.~(\ref{OmegaHSchroedinger}) from
Eq.~(\ref{HOmegaSchroedinger}) leads to
\beq
\big[\Omega, H_0\big]\eta |\Psi\rangle = \big(V - \Omega \eta
V\big)|\Psi\rangle\nn
=\big(V - \Omega \eta
V\big)\Omega \eta|\Psi\rangle\,.
\eeq
This way we obtain a nonlinear equation for the M{\o}ller operator $\Omega$
\beq
\big[\Omega,H_0\big]-V \Omega  + \Omega V \Omega =0.\label{Omega:rel}
\eeq
Defining the operator $A$ via $\Omega =: \eta + A$ with $A\,=\,\lambda
A \eta$, as follows from
Eqs.~(\ref{moeller:eta:right})
and~(\ref{moeller:eta}), we rewrite Eq.~(\ref{Omega:rel}) in the form 
\beq
\lambda\big(H +\big[H,A\big] - A
V A \big) \eta = 0.\label{decoupling:eq:A}
\eeq
The UT $U$ was parametrized by Okubo ~\cite{Okubo:1954zz} in terms of
the operator $A$
via
\beqa
U&=&\left(
\begin{array}{rr}
\eta\,(1+A^\dagger A)^{-1/2} & -A^\dagger(1+A A^\dagger)^{-1/2} \\
A (1+A^\dagger A)^{-1/2}& \lambda (1+A A^\dagger)^{-1/2}\\
\end{array}
\right).\label{OkuboTransf}
\eeqa
The resulting transformed Hamiltonian 
\beqa
\eta \,U^\dagger H \,U \eta&=&(\Omega^\dagger\Omega)^{1/2}\eta H \,\Omega
(\Omega^\dagger \Omega)^{-1/2} ,\label{effectivePotentialUT}
\eeqa
leads to the effective potential defined via
\beq
V_{\rm eff}^{\rm UT} :=  \eta \,U^\dagger H \,U \eta - H_0.
\eeq
Obviously, the Okubo transformation in Eq.~(\ref{OkuboTransf}) is
not the only possibility to obtain a block-diagonalized Hamiltonian. On
top of the transformation $U$ one can always apply e.g.~a UT acting
nontrivially on the $\eta$-space, thus leaving the Hamiltonian block-diagonal.
This freedom 
has been exploited in a systematic manner
to construct renormalizable/factorizable 3NFs and four-nucleon
forces (4NFs) in chiral EFT in Refs.~\cite{Epelbaum:2006eu,Bernard:2007sp,Bernard:2011zr,Krebs:2012yv,Krebs:2013kha}.

To derive the potential $V_{\rm eff}^{\rm UT}$ from the effective
chiral Lagrangian in Eq.~(\ref{Lagr}) one
needs to solve the nonlinear decoupling equation~(\ref{decoupling:eq:A})
for the operator $A$. This can be done perturbatively using
NDA \cite{Epelbaum:2005pn} to count 
powers of three-momenta and pion masses, denoted collectively by $Q$.
For the sake of definiteness, we restrict ourselves in the following to nuclear
potentials in the absence of external sources. The extension to the
current operators is straightforward and discussed in details in
Ref.~\cite{Krebs:2016rqz}. 
The irreducible contributions of any connected Feynman diagram scale as $Q^\nu$
with $\nu=-2 + \sum_i V_i\kappa_i$, where 
$V_i$ denotes the number of vertices of type $i$ and $\kappa_i$
is the inverse mass dimension of the corresponding coupling constant,
$\kappa_i =  d_i + \frac{3}{2}n_i + p_i - 4$.
Here, $d_i$ is the number of derivatives and/or $M_\pi$-insertions, while  $n_i$ and
$p_i$  denote the number of nucleon 
and pion fields,
respectively.\footnote{Alternatively (but equivalently), the chiral
  order $\nu$ of a connected, $N$-nucleon irreducible diagram with $L$ loops can
  be expressed as $\nu=-4 + 2 N + 2 L + \sum_i V_i\Delta_i$ with
  $\Delta_i = d_i + n_i/2 - 2$.}
This particular form of the power counting allows one to formulate the
chiral expansion in the form that is completely analogous to the expansion in
powers of coupling constants. It is thus particularly well suited for
algebraic approaches such as the method of UT. 
Once the operator $A$ is
available, one can perform the chiral expansion of
Eq.~(\ref{effectivePotentialUT})  to construct the effective
potential order-by-order.

\begin{figure}[tb]
\begin{center}
\includegraphics[width=1.0\textwidth,angle=0,trim = 20mm 70mm 00mm 00mm]{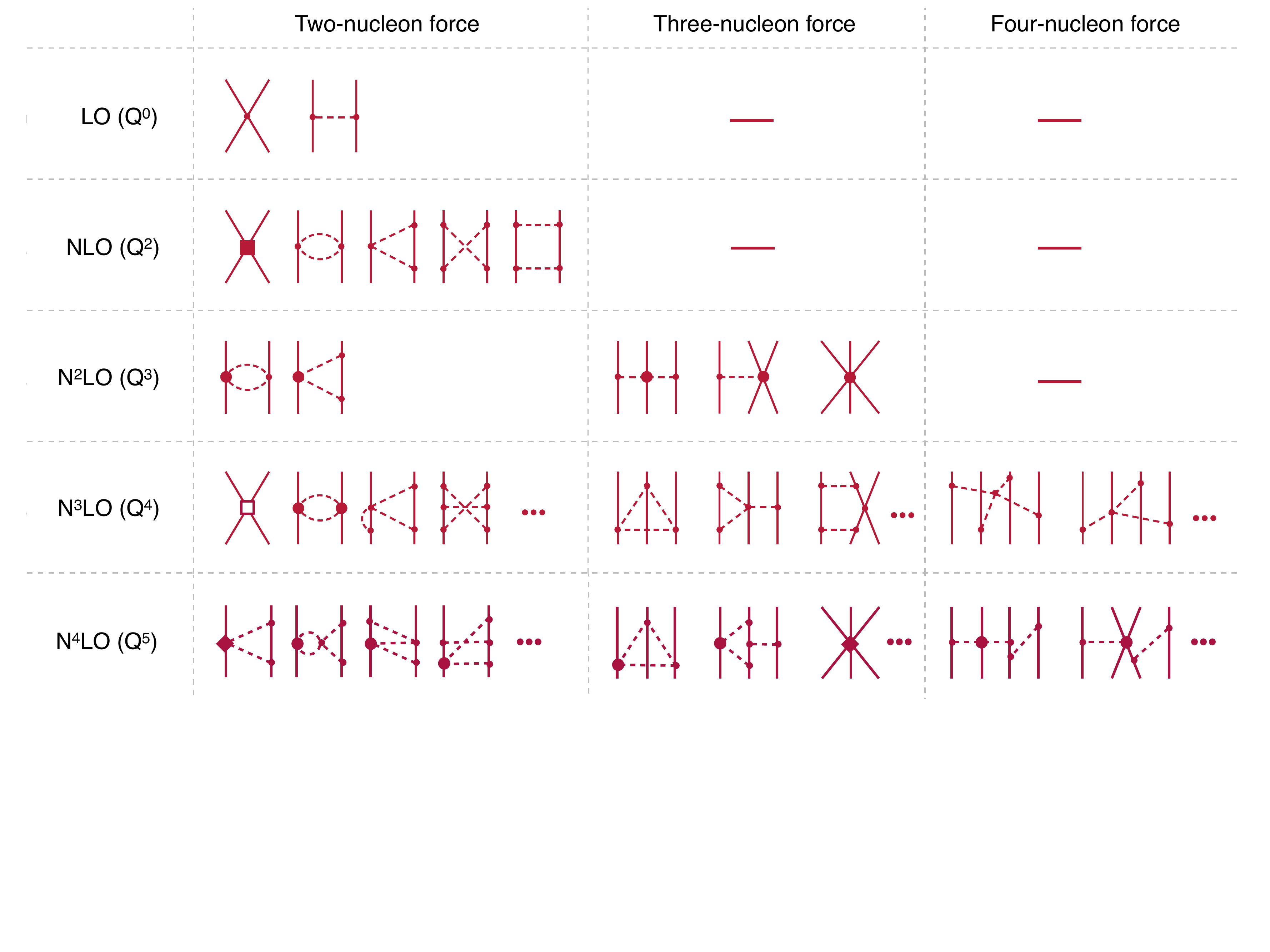}
\end{center}
\caption{Hierarchy of nuclear forces at 
  increasing orders in chiral expansion in the Weinberg scheme. Solid
  and dashed lines refer to nucleons and pions, respectively. Solid
  dots, filled circles, filled squares, filled
  diamonds  and open squares refer to vertices from the Lagrangian in Eq.~(\ref{Lagr})
  of dimension $\Delta = 0$, $1$, $2$, $3$ and $4$, respectively.}\label{chiralexpansionNN}
\end{figure}

The chiral expansion of the nuclear forces is visualized in
Fig.~\ref{chiralexpansionNN}. Below, we briefly discuss 
isospin symmetric contributions starting from the leading order (LO) $Q^0$.
The only contributions at this order emerge from the 
OPEP and two contact interactions $\propto C_{S,T}$
\cite{Weinberg:1990rz,Weinberg:1991um}. The first corrections at order
$Q^2$ (NLO) involve  the leading two-pion exchange potential (TPEP)
\cite{Ordonez:1995rz,Kaiser:1997mw,Epelbaum:1998ka} and
$7$ short range interactions $\propto C_i$.
At order $Q^3$ (N$^2$LO), further corrections to the TPEP
$\propto c_i$ need to be taken into account \cite{Kaiser:1997mw}.
At the same order one has the first nonvanishing contributions to the 3NF. They are given by
the two-pion exchange diagram involving the LECs $c_i$ and two
shorter-range tree-level diagrams involving the LECs $D$ and $E$
\cite{vanKolck:1994yi,Epelbaum:2002vt}.
At order $Q^4$ (N$^3$LO), the NN potential receives the contributions
from the
leading three-pion exchange 
\cite{Kaiser:1999jg,Kaiser:1999ff,Kaiser:2001dm}, further corrections
to the TPEP \cite{Kaiser:2001pc,Kaiser:2001at} and 12 new short-range
interactions $\propto D_{i}$ \cite{Reinert:2017usi}. At  the same order, there are various
one-loop corrections to the 3NF \cite{Bernard:2007sp,Bernard:2011zr,Ishikawa:2007zz} and the first contributions to the
4NFs \cite{Epelbaum:2006eu,Epelbaum:2007us}, which do not involve
unknown parameters. Finally, at order $Q^5$
(N$^4$LO), the NN potential receives corrections to
the three-pion exchange
$\propto c_i$ \cite{Kaiser:2001dm} and further contributions to 
the TPEP \cite{Entem:2014msa}.   
No additional unknown parameters appear in the isospin-conserving part
of the NN force at this order.
The 3NF also receives corrections at N$^4$LO, most of which
have already been worked out using DR \cite{Krebs:2012yv,Krebs:2013kha,Girlanda:2011fh}. Notice that the 3NF
involves at this order a number of new short-range operators.
Work is still in progress to
derive the remaining 3NF and 4NF at N$^4$LO. We further emphasize that
all calculations
mentioned above are carried out using DR or equivalent schemes.

The effective potential $V_{\rm eff}^{\rm UT}$ leads, by construction, to the same spectrum
and on-shell scattering matrix as the original untransformed potential
$V$~\cite{Epelbaum:1998na,Polyzou:2010eq}. There are, however, other
possibilities to define the effective potential without
changing on-shell physics. One example is an
energy-independent potential defined by
\beqa
V_{\rm eff}^{\rm EI}&=&\eta V \,\Omega\,\eta.\label{energyindepVeff}
\eeqa
The proof that $V_{\rm eff}^{\rm EI}$ of Eq.~(\ref{energyindepVeff})
leads to the same spectrum is trivial:
\beqa
\big(\eta\, H_0\,\eta + \eta \,V\,\eta  + \eta
\,V\,\lambda\big)|\Psi\rangle&=&E\, \eta|\Psi\rangle,\nn
\big(\eta\, H_0\,\eta + \eta \,V\,\eta  + \eta
\,V\,\lambda\,\Omega\big)\eta|\Psi\rangle&=&E\, \eta|\Psi\rangle,\nn
\big(\eta\, H_0\,\eta + \eta \,V\,\Omega \eta\big)\eta|\Psi\rangle&=&E\, \eta|\Psi\rangle,
\eeqa
where we used Eqs.~(\ref{moeller:op:def}) and (\ref{moeller:eta}) in
the first and second lines, respectively. Note that the potential $V_{\rm
  eff}^{\rm EI}$ is manifestly non-hermitian. However, due to its
simplicity, it is widely used in the
literature~\cite{Suzuki:PTP1983}. This example shows that 
there is a considerable freedom to define nuclear potentials. Nuclear
forces and current operators constructed by the Bochum-Bonn group, see
e.g.~Refs.~\cite{Epelbaum:1998ka,Epelbaum:2007us,Bernard:2007sp,Bernard:2011zr,Kolling:2011mt,Krebs:2012yv,Krebs:2013kha,Epelbaum:2014efa,Epelbaum:2014sza,Krebs:2016rqz,Reinert:2017usi},
are obtained using the method of UT. The JLab-Pisa group utilizes a different approach
by 
starting with the on-shell transfer matrix $T$ and
``inverting'' it to obtain the effective potential, see
e.g.~Refs.~\cite{Pastore:2009is,Pastore:2011ip,Piarulli:2012bn,Baroni:2015uza}.
This is carried out
in perturbation theory by counting the nucleon mass via $m\sim
\Lambda_{\rm b}$
\beq
T=T^{(0)} + T^{(1)} + T^{(2)} + \dots\;,
\eeq
where the superscripts indicate the chiral order $Q^n$. The same
counting scheme is used to organize the contributions
to effective potential:
\beq
v=v^{(0)} + v^{(1)} +v^{(2)} + \dots\;.
\eeq
The inversion of the LS equation is carried out iteratively to
yield
\beq
v^{(0)} = T^{(0)}\,,\quad \quad  
v^{(1)} =  T^{(1)} - v^{(0)} G_0 v^{(0)} \,, \quad \quad
\ldots .\label{inversion:Tmatrix}
\eeq
Obviously, the knowledge of the on-shell transfer matrix is insufficient
to perform the inversion, and one needs to specify its off-shell
extension.
Notice that the potentials constructed in this way are not necessarily
hermitian, and thus there is no guarantee that they are unitarily equivalent
to the ones derived using the UT technique. It should, however, always be
possible to find a similarity transformation that 
relates one potential to another. This is exemplified
with the potential $v_{2\pi}^{(3)}(\nu=1)$ in
Eq.~(20) of Ref.~\cite{Pastore:2011ip}, where $\nu$ is an arbitrary
phase,
which is manifestly non-hermitian. Using the similarity
transformation in Eq.~(28) of that paper\footnote{The claim in
  Ref.~\cite{Pastore:2011ip} that the transformation $e^{i U}$ in Eq.~(28)
  of that paper is unitary  is incorrect since the operator $i U^{(1)}(\nu)$
from Eq.~(28) is not antihermitian.}, it can be transformed to 
the hermitian potential $v_{2\pi}^{(3)}(\nu=0)$, that is actually
employed in the current version of
the interactions developed by the JLab-Pisa group. With this choice, their potentials are 
unitarily equivalent to the ones of the Bochum-Bonn group.

\section{Regularization}
\label{sec4}

\subsection{Semilocal momentum-space regularization of the NN potential}
\label{sec41}

In this review article we focus on the semilocal regularization
approach of the chiral
nuclear potentials carried out in momentum space \cite{Reinert:2017usi}. For the purpose
of regularization we will consider the two-nucleon interaction
consisting of two distinct parts: the short-range contact interaction
part and the long-range pion-exchange part. In this context, the term
"semilocal" refers to the application of a nonlocal regulator for the
former and a local regulator for the latter. In particular, the
momentum-space matrix elements of the contact potential are multiplied
by a simple nonlocal Gaussian regulator 
\begin{equation}
	\label{eq:contact-interaction-regularization}
    \langle \vec p\, ' | V_{\rm{cont}} | \vec p \, \rangle_{\rm reg} =
    \langle \vec p \, ' | V_{\rm cont} | \vec p \, \rangle \; e^{-\frac{p'^2+p^2}{\Lambda^2}}.
\end{equation}
Here and in what follows, $p \equiv | \vec p \, |$ and $p'  \equiv | \vec
p \, ' |$. 
Such kinds of nonlocal regulators
(albeit with
different powers of $p$, $p'$ and $\Lambda$) have been and still are
employed as the main method of regularization for the entire
potential including the long-range interactions, see
e.g.~Refs.~\cite{Entem:2003ft,Epelbaum:2004fk,Carlsson:2015vda,Ekstrom:2015rta,Entem:2017gor,Ekstrom:2017koy}. \footnote{Notice   
  that the aforementioned potentials (except the one of
  Ref.~\cite{Entem:2003ft}) additionally apply spectral function
  regularization (SFR) \cite{Epelbaum:2003gr,Epelbaum:2003xx} of the
  TPEP in the
  form of a sharply cut-off spectral integral in order to
  suppress its remaining unphysical short-distance behavior. Notice, however,
  that the application of a nonlocal regulator
  $\exp(-(p^{2n}+p'^{2n})/\Lambda^{2n})$ with suitably chosen $n$ is
  sufficient to arrive at UV-finite iterations of the potential.} 

However, in Refs.~\cite{Epelbaum:2014efa,Epelbaum:2014sza} it was
shown that the amount of long-range cutoff artifacts can be
significantly reduced by employing a local regulator for
pion-exchange potentials. Notice that pion-exchange contributions,
except for some relativistic corrections, give rise to local
potentials.
We 
\emph{require} the regulator to preserve the analytic structure of the
scattering amplitude and, in
particular, the left-hand cuts in the complex energy plane related to
the long-range part of the interaction, which are 
unambiguously determined in chiral
EFT. 
Inspired by Ref.~\cite{Rijken:1990qs}, this is achieved in our
momentum-space approach by regularizing the static propagators
of pions exchanged between different nucleons 
with a local Gaussian cutoff via  
\begin{equation}
	\label{eq:pion-propagator-form-factor}
    \frac{1}{l^2+M_\pi^2} \rightarrow \frac{1}{l^2+M_\pi^2} \; e^{-\frac{l^2+M_\pi^2}{\Lambda^2}},
\end{equation}
with $l = |\vec{l}|$ and $\vec{l}$ denoting the three-momentum of the
exchanged pion. The introduction of the Gaussian form factor in the
pion propagators leads to properly regularized long-range potentials
that are finite at short distances in coordinate space. In order to
have a clean separation of the long-range pion-exchange potential from
the short-range contact interactions, we made use of the available
contact interactions to subtract out the remaining (finite)
admixtures of 
short-range interactions \cite{Reinert:2017usi}. The fixed coefficients of these
subtractions are determined from the requirement that the
corresponding coordinate-space potential and as many derivatives
thereof as allowed by power counting vanish at the origin. This
convention leads to a qualitatively similar regularization as the
coordinate-space regulator previously employed in
Refs.~\cite{Epelbaum:2014efa,Epelbaum:2014sza}. 

Application of these ideas to the OPEP is straightforward and leads,
in the limit of exact isospin
symmetry,
to 
\begin{equation}
    \label{eq:OPEP-regularized}
    V^{1\pi}_{\rm 2N, \, \Lambda} (M_\pi) = -\frac{g_A^2}{4F_\pi^2} \, \fet \tau_1
    \cdot \fet \tau_2  \left(
      \frac{\vec \sigma_1  \cdot \vec q \, \vec \sigma_2 \cdot \vec
        q}{q^2+M_\pi^2} + C(M_\pi) \, \vec \sigma_1 \cdot \vec
      \sigma_2 \right) e^{-\frac{q^2 + M_\pi^2}{\Lambda^2}}, 
\end{equation}
where $q \equiv | \vec q \,| \equiv  | \vec p\, ' - \vec p \,|$ and  $\vec \sigma_i$  ($\fet \tau_i$) are the Pauli
spin (isospin) matrices of the $i$-th nucleon.  
Here, the static pion propagator has been regularized according to
Eq.~\eqref{eq:pion-propagator-form-factor} and a likewise-regularized
LO contact interaction has been added to the OPEP. Its coefficient
$C(M_\pi)$,
\begin{equation}
	C(M_\pi) = -\frac{\Lambda \left(\Lambda ^2-2 M_\pi^2 \right) + 2 \sqrt{\pi } M_\pi^3 e^{\frac{M_\pi^2}{\Lambda ^2}}
   \text{erfc}\left(\frac{M_\pi}{\Lambda }\right)}{3 \Lambda ^3} \,,
\end{equation}
with $\text{erfc} (z)$ denoting the complementary error function,  
is fixed by the requirement that the spin-spin part of the
OPEP in coordinate space vanishes at the origin. 
For the regularization of the TPEP, we start with a generic
three-dimensional loop integral $I(\vec q \,)$ arising in the
derivation of the TPEP using e.g.~the method of unitary transformation as
detailed in the previous section or comparable approaches like
time-ordered perturbation theory or S-matrix-based methods
\cite{Kaiser:1997mw}. As discussed in Ref.~\cite{Rijken:1990qs}, the
pion energy denominators in the corresponding $1$-loop expressions can always
be rewritten into an integral over a mass parameter $\lambda$
involving a product of two static pion propagators with mass
$\sqrt{M_\pi^2+\lambda^2}$ 
\begin{equation}
	\label{eq:TPEP-loop-integral}
	I(\vec q \,) = \int_0^\infty d \lambda \, \int \frac{d^3 l_1}{(2 \pi)^3} \, \frac{d^3 l_2}{(2 \pi)^3} \, (2 \pi)^3 \delta (\vec q - \vec l_1 - \vec l_2 ) \, \frac{1}{(l_1^2 + M_\pi^2 + \lambda^2 ) (l_2^2 + M_\pi^2 + \lambda^2)} \times \ldots \,,
\end{equation}
where $\vec l_1$ and $\vec l_2$ denote the three-momenta of the
exchanged pions and the ellipses refer to additional
momentum-spin-isospin structures arising from the vertices of a
particular diagram. With the pion propagators factorized in this a
way, we can regularize them by applying the prescription
specified in  Eq.~\eqref{eq:pion-propagator-form-factor} to each of them. Although
the introduction of the regulator obviously affects the resulting
expression for the TPEP, there is no need to rederive them
explicitly. Indeed, the scalar functions accompanying the spin-isospin
operators in the \emph{unregularized} TPEP can be expressed using  the
dispersive representation 
\begin{equation}
    \label{eq:spectral-integral}
    V^{2\pi}_{\rm 2N}  (q) = \frac{2}{\pi}\int_{2M_\pi}^\infty \mu \, d\mu \frac{\rho(\mu)}{q^2+\mu^2} \, ,
\end{equation}
 with the spectral functions $\rho(\mu) =
 \Im(V_{2\pi}(q))\vert_{q=0^+-i\mu}$ which are readily available up to
 N$^4$LO. For the explicit expressions of the TPEP, additional
 subtractions of short-range terms have to be performed to arrive at a
 convergent spectral integral in Eq.~\eqref{eq:spectral-integral}
 whose number depends on the chiral
 order of the contribution at hand. Introducing the pion propagator
 regulators in Eq.~\eqref{eq:TPEP-loop-integral}, the regularized
 generic spectral integral of Eq.~\eqref{eq:spectral-integral} takes
 the form 
\begin{equation}
    \label{eq:TPEP-regularized}
    V^{2\pi}_{{\rm 2N}, \Lambda}(q) = e^{-\frac{q^2}{2\Lambda^2}} \, \frac{2}{\pi}\int_{2M_\pi}^\infty \mu \, d\mu \frac{\rho(\mu)}{q^2+\mu^2} \, e^{-\frac{\mu^2}{2\Lambda^2}} \, .
\end{equation}
As the spectral  representation in
Eq.~\eqref{eq:spectral-integral} is a dispersion relation, it makes
the analytic structure of the TPEP as a function of the
momentum-transfer $\vec q$ manifest. Indeed, one could immediately
infer the form of the regulator as a function of $q^2$ and $\mu^2$ by
the requirement to preserve the analytic structure of the TPEP.
The additional factor of $1/2$ in the exponent of
Eq.~\eqref{eq:TPEP-regularized} would, however, not have been taken into account
correctly if we had naively applied our prescription
Eq.~\eqref{eq:pion-propagator-form-factor} to the propagator-like
denominator of the spectral integral representation. 

Expanding the exponentials in inverse powers of the cutoff in
either Eq.~\eqref{eq:OPEP-regularized} or
Eq.~\eqref{eq:TPEP-regularized}, one observes that the regulator
indeed does not affect the long-range part of the potential to any
order, but generates an infinite
series of short-range terms polynomial in $q^2$.
Since an increasing number of 
contact interactions of this form with freely adjustable LECs
become available with increasing chiral order, the perturbative
restoration of cutoff-independence is also obvious in this scheme. 

The expressions of the regularized and subtracted TPEP can be found in
Ref.~\cite{Reinert:2017usi}. Here we restrict ourselves to the example of
the isospin-independent central part of the leading TPEP at NLO which is given by 
\begin{equation}
    \label{eq:WC-NLO}
    W_{C, \, \Lambda}^{(2)} (q) = e^{-\frac{q^2}{2 \Lambda^2}} \;
    \frac{2}{\pi} \int_{2 M_\pi}^\infty \, \frac{d \mu}{\mu^3}
    \rho_{C}^{(2)} (\mu) \bigg(\frac{q^4} {\mu^2 + q^2} + C_{C, 1}^2
    (\mu ) + C_{C, 2}^2 (\mu ) \, q^2\bigg) \, e^{- \frac{\mu^2}{2
        \Lambda^2}} \,,
  \end{equation}
  with the spectral function
  \beq
 \rho_{C}^{(2)} (\mu) = \frac{\sqrt{\mu^2 - 4 M_\pi^2}}{768 \pi
   F_\pi^4 \mu}
\bigg(4M_\pi^2 (5g_A^4 - 4g_A^2 -1)  - \mu^2 (23g_A^4 - 10g_A^2 -1)
+ \frac{48 g_A^4 M_\pi^4}{4 M_\pi^2 -\mu^2} \bigg)\,.
  \eeq
Two subtractions have been performed in order to render the
unregularized spectral integral in Eq.~\eqref{eq:WC-NLO} convergent
and according to our convention, we have additionally fixed the
subtraction coefficients $C_{C, 1}^2 (\mu )$ and $C_{C, 2}^2 (\mu )$
by the requirement that $W_{C, \, \Lambda}^{(2)} (r) \Big \vert_{r=0} =
\frac{d^2}{dr^2} W_{C, \, \Lambda}^{(2)} (r) \Big \vert_{r=0} = 0$. (The
first derivative of $W_{C, \, \Lambda}^{(2)} (r)$ vanishes at the
origin regardless of the subtraction coefficients.) 
\begin{figure}[tb]
    \begin{center}
        \includegraphics[width=0.5\textwidth,keepaspectratio,angle=0,clip]{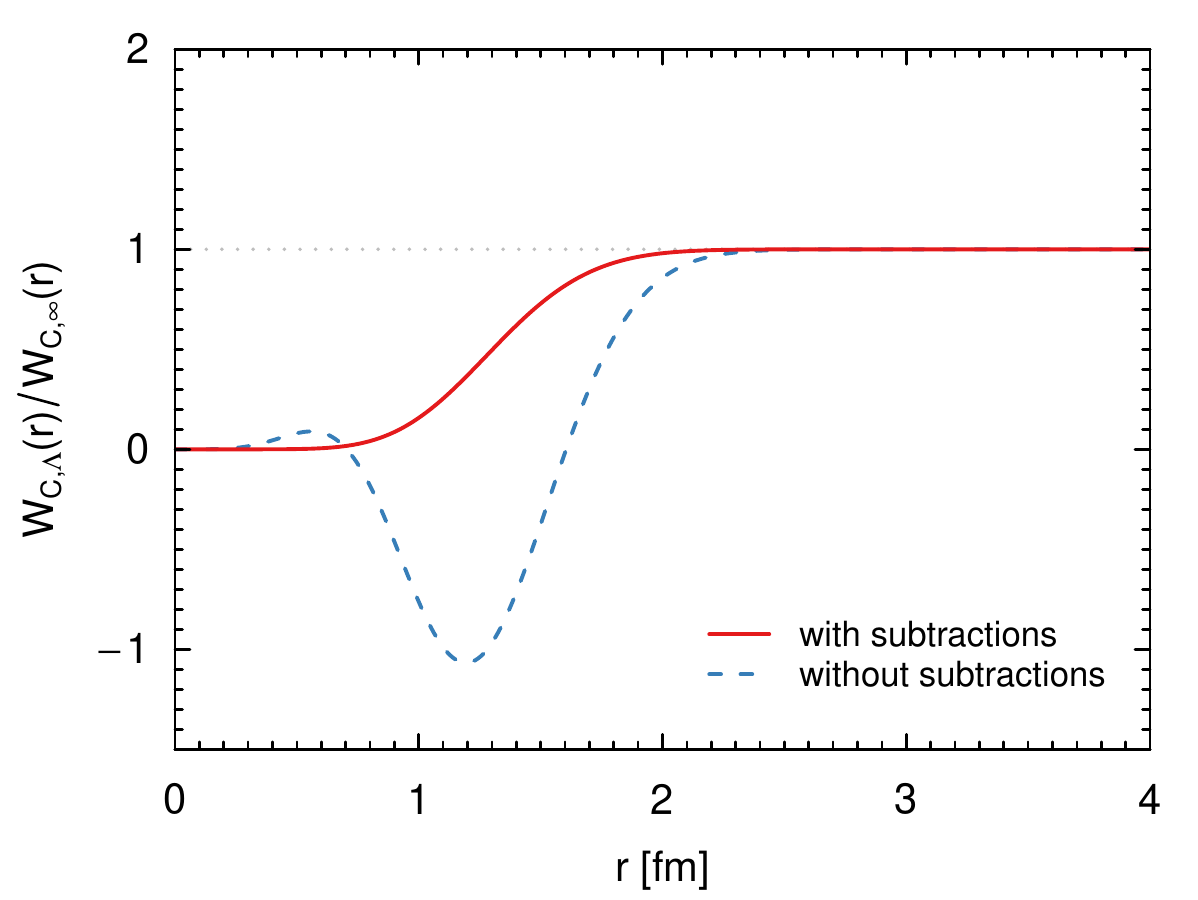}
    \end{center}
    \caption{  
        \label{fig:subtractions-WC-NLO}
        Ratio of the regularized and unregularized central part of the leading TPEP in coordinate space for $C_{C, 1}^2 (\mu )$, $C_{C, 2}^2 (\mu )$ fixed as discussed in the text and $C_{C, 1}^2 (\mu ) = C_{C, 2}^2 (\mu ) = 0$.
    }
\end{figure}
Figure \ref{fig:subtractions-WC-NLO} shows the ratio of the regularized
and unregularized expressions in Eq.~\eqref{eq:WC-NLO} in coordinate
space. As one can see, the behavior of the regularized potential is
smoother when fixing the subtraction coefficients by the convention
explained above. Also note that the potential with $C_{C, 1}^2 (\mu )
= C_{C, 2}^2 (\mu ) = 0$ does not vanish at the origin.\footnote{This
  is not visible in Fig.~\ref{fig:subtractions-WC-NLO} since the
  unregularized potential $W_{C, \infty} (r)$ is singular at $r = 0$.}

\subsection{Regularization and consistency of nuclear forces}
\label{subsec:Reg}

Having defined the regularization scheme in the NN sector, we now turn
to regularization of the 3NF. 
The expressions for the 3NFs described in section~\ref{sec3} have been worked out
completely  through N$^3$LO using DR. They are off-shell consistent with the
unregularized NN interactions reviewed in that section in the way
explained in section~\ref{sec2}. To arrive at regularized 3NFs, it is
tempting to apply some kind of multiplicative regulators to the
expressions of the 3NF derived using DR.
Such a naive approach, however, leads to a violation
of the chiral symmetry at N$^3$LO and destroys the consistency between
two- and three-nucleon forces after regularization.

To illustrate the problem consider the diagrams shown in Fig.~\ref{fig:1}, which have already
been discussed in section~\ref{sec2}.
The 3NF entering the first graph on the right-hand side (r.h.s.) is
given by \cite{Bernard:2011zr}
\beq
V_{\rm 3N}^{2\pi, \, 1/m} = i\frac{g_A^2}{32 m F_\pi^4}\frac{\vec{\sigma}_1\cdot\vec{q}_1\,\vec{\sigma}_3\cdot\vec{q}_3}{(q_1^2+M_\pi^2)(q_3^2+M_\pi^2)}\fet{\tau}_1\cdot(\fet{\tau}_2\times\fet{\tau}_3)(2\vec{k}_1\cdot\vec{q}_3+4\vec{k}_3\cdot\vec{q}_3+i\,[\vec{q}_1\times\vec{q}_3]\cdot\vec{\sigma}_2)\,,\quad\quad\label{oneovermV2pigA2}
\eeq
with $\vec{q}_i = \vec{p}_i^{\,\prime}-\vec{p}_i$, $\vec{k}_i= 1/2 \big(\vec{p}_i^{\,\prime}-\vec{p}_i\big)$
and $\vec{p}_i$, ($\vec{p}_i^{\,\prime}$) the initial (final)
momenta of the $i$-th nucleon. The complete expression for the relativistic
corrections to the 3NF at N$^3$LO can be found 
in~Ref.~\cite{Bernard:2011zr}. We now consider the first iteration of
$V_{\rm 3N}^{2\pi, \, 1/m}$ with the static OPEP
\beq
 V_{\rm 2N}^{1\pi }= - \left(\frac{g_A}{2 F_\pi}\right)^2\fet{\tau}_1\cdot\fet{\tau}_2\frac{\vec{\sigma}_1\cdot\vec{q}\,\vec{\sigma}_2\cdot\vec{q}}{q^2 + M_\pi^2}
\eeq
as shown by the first diagram on the r.h.s.~of Fig.~\ref{fig:1}. By
simply counting the powers of momenta in the 
loop integration one observes that the loop integral is linearly
divergent, which leads to a finite result in DR. As already pointed
out in section~\ref{sec2}, adding the  DR expression for the
3NF $V_{\rm 3N}^{2\pi-1\pi}$ from Eqs.~(2.16)-(2.20) of
Ref.~\cite{Bernard:2007sp} yields (on-shell) the same result 
as obtained from calculating the Feynman diagram on the l.h.s.~of
Fig.~\ref{fig:1} as expected for consistent two- and three-nucleon
forces.

We now repeat this exercise using the
semilocally regularized nuclear potentials
\beq
V_{\rm 3N, \, \Lambda}^{2\pi, \, 1/m} \, = \, V_{\rm 3N}^{2\pi,
  \, 1/m} \; e^{-\frac{q_1^2+M_\pi^2}{\Lambda^2}}\, 
e^{-\frac{q_3^2+M_\pi^2}{\Lambda^2}},
\quad \quad
V_{\rm 2N, \, \Lambda}^{1\pi }\,=\, V_{\rm 2N}^{1\pi } \; e^{-\frac{q^2+M_\pi^2}{\Lambda^2}} \,,
\eeq
in the calculation of the
first diagram on the r.h.s.~of
Fig.~\ref{fig:1}. This leads  to
\beqa
V_{\rm 3N, \, \Lambda}^{2\pi, \, 1/m} \, G_0  \, V_{\rm 2N, \,
  \Lambda}^{1\pi } \; + \; V_{\rm 2N, \,
  \Lambda}^{1\pi }  \, G_0  \, V_{\rm 3N, \, \Lambda}^{2\pi, \, 1/m}  
& =& 
\Lambda\frac{g_A^4}{128\sqrt{2}\pi^{3/2}F_\pi^6}(\fet \tau_2\cdot \fet
\tau_3-\fet \tau_1\cdot \fet \tau_3)
\frac{{\vec{q}_2}\cdot\vec{\sigma}_2
  \vec{q}_3\cdot\vec{\sigma}_3}{q_3^2+M_\pi^2} \nn
&-&\Lambda\frac{g_A^4}{96\sqrt{2}\pi^{3/2}F_\pi^6}\frac{\vec{q}_3\cdot\vec{\sigma}_3
  \vec{q}_3\cdot\vec{\sigma}_1 \; \fet{\tau}_1\cdot
  \fet{\tau}_3}{q_3^2+M_\pi^2}+\dots\;,\label{chiralsymmetryviolatingDivergence} 
\eeqa
where the ellipses refer to all permutations of the nucleon labels and
terms finite in the $\Lambda \to \infty$-limit. The linear divergence
$\propto \vec{q}_3\cdot\vec{\sigma}_3
  \vec{q}_3\cdot\vec{\sigma}_1$ is cancelled by the $D$ counter term in the second
3NF diagram at N$^2$LO  in Fig.~\ref{chiralexpansionNN}. To cancel the
linearly divergent contribution $\propto 
{\vec{q}_2}\cdot\vec{\sigma}_2$ one would, however,  need to introduce
a vertex in $\mathcal{L}_{\pi \rm NN}^{(1)}$
corresponding to a derivative-less coupling of the pion to the NN systems.
Such vertices violate the chiral symmetry and, being  suppressed by
powers of $M_\pi^2$, cannot appear in
$\mathcal{L}_{\pi \rm NN}^{(1)}$.  As a consequence, this linear
divergence can \emph{not}
be absorbed into redefinition of the LECs, and the amplitude on the
r.h.s.~of Fig.~\ref{fig:1} can seemingly not be renormalized (i.e.~made finite in the $\Lambda \to
\infty$ limit). The r.h.s.~of the shown equation, therefore, apparently
cannot match the (renormalizable) on-shell scattering amplitude from the Feynman
diagram on the l.h.s.. The problem
can be traced back to mixing the DR when calculating the 3NF $V_{\rm
  3N}^{2\pi-1\pi}$ with a cutoff regularization for the
iterative contributions in
Eq.~(\ref{chiralsymmetryviolatingDivergence}), see Ref.~\cite{Krebs:2019uvm} for
another example with the NN axial vector current operator at N$^3$LO.   
Indeed, recalculating the loop integral
in
$V_{\rm 3N}^{2\pi-1\pi}$ using the cutoff-regularized
pion propagators leads to 
\beq
V _{\rm 3N,  \, \Lambda}^{2\pi-1\pi} = -\Lambda\frac{g_A^4}{128\sqrt{2}\pi^{3/2}F_\pi^6}(\fet
\tau_2\cdot \fet \tau_3-\fet \tau_1\cdot \fet \tau_3)
\frac{{\vec{q}_2}\cdot\vec{\sigma}_2
  \vec{q}_3\cdot\vec{\sigma}_3}{q_3^2+M_\pi^2}
-\Lambda\frac{g_A^4}{32\sqrt{2}\pi^{3/2}F_\pi^6}\frac{\vec{q}_3\cdot\vec{\sigma}_3
  \vec{q}_3\cdot\vec{\sigma}_1 \; \fet{\tau}_1\cdot \fet{\tau}_3}{q_3^2+M_\pi^2}+\dots\;,\label{twopiononepionDivergence}
\eeq 
where the ellipses refer to the finite terms in the $\Lambda \to \infty$-limit. The problematic linear
divergence cancels exactly and the agreement with the 
on-shell amplitude from the Feynman diagram is restored 
when both \emph{consistently regularized}
contributions on the r.h.s.~of Fig.~\ref{fig:1} are added together.

One may worry whether the regularization issues discussed above could also be 
relevant for NN interactions. Fortunately, this is not the
case since the momentum structure of the NN contact interactions is
not restricted by the chiral symmetry. UV divergences emerging
from iterations of the LS equation can, therefore, always be absorbed
into redefinition of the bare LECs $C_{S, T} (\Lambda )$, $C_{i} (\Lambda )$, $\ldots$.

In the considered example with the 3N amplitude, the consistently regularized 3NF could be
obtained by simply recalculating $V_{\rm 3N}^{2\pi-1\pi}$ with all pion
propagators being
regularized according to Eq.~(\ref{eq:pion-propagator-form-factor}). 
This would indeed solve the problem with the cancelation of linear
divergencies at N$^3$LO, but it would still lead to a violation of the chiral
symmetry in diagrams involving three- and four-pion vertices, which 
depend on the parametrization of the pion field. For vertices involving
up to four pion fields, this freedom is represented 
by a single real parameter $\alpha$. In the effective chiral
Lagrangian, all pion fields are collected in an SU(2) matrix
$U(\fet\pi)$, whose most general expression, expanded in powers of the
pion fields, takes the form
\beqa
U(\fet\pi)&=&1+\frac{i}{F} \fet{\pi}\cdot\fet{\tau}
-\frac{1}{2 F^2}
(\fet{\pi}\cdot\fet{\tau})^2-\alpha\frac{i}{F^3}(\fet{\pi}\cdot\fet{\tau})^3
+
\bigg(\alpha-\frac{1}{8}\bigg)\frac{1}{F^4}(\fet{\pi}\cdot\fet{\tau})^4
+ {O}({\fet \pi}^5)\,. \label{alphaParametrizationU}
\eeqa
Clearly, the on-shell
amplitude must be independent of the arbitrary parameter
$\alpha$. Evaluating the 3NF and 4NF with the regularized pion
propagators, however, leads to $\alpha$-dependent expressions (for finite
values of $\Lambda$). This shows, perhaps not surprisingly, that the simplistic
approach by regularizing all pion propagators as described
above violates the chiral  symmetry. A possible
solution of this problem is provided by the symmetry preserving higher derivative
regularization method introduced by Slavnov \cite{Slavnov:1971aw}, see
also~Refs.~\cite{Djukanovic:2004px,Long:2016vnq} for recent applications in chiral EFT.

To summarize, we have shown that a naive regularization of the three- and
more-nucleon
forces by multiplying the
expressions
derived in DR with regulator functions leads to inconsistencies  
starting from N$^3$LO, see Ref.~\cite{Krebs:2019uvm} for the same conclusion for 
two- and more-nucleon charge and
current operators. 
This problem is by no means
restricted to
semilocal cutoffs.
To derive many-body forces and
currents regularized \emph{consistently} with the NN potentials of Ref.~\cite{Reinert:2017usi}, the
expressions for the 3NF of Refs.~\cite{Bernard:2007sp,Bernard:2011zr,Krebs:2012yv}, 4NF of Ref.~\cite{Epelbaum:2007us} and exchange
charge and current operators of
Refs.~\cite{Kolling:2009iq,Kolling:2011mt,Krebs:2016rqz,Krebs:2019aka}
need to be
recalculated using e.g.~an appropriately chosen higher derivative regulator at the level of the
effective Lagrangian.

\section{Truncation error analysis}
\label{sec5}

Estimating the uncertainty associated with truncations of the EFT
expansion, which typically dominates the error budget (see section
\ref{sec6}), is an important task -- in particular since chiral EFT is
being developed into a precision tool. In the past, truncation errors were often
estimated in few-nucleon calculations from a residual cutoff
dependence. This approach, however, suffers from several drawbacks 
and does not allow for a reliable estimation of truncation errors
\cite{Epelbaum:2004fk}.  In Ref.~\cite{Epelbaum:2014efa}, we have formulated a simple 
algorithm to estimate the size of neglected higher-order terms
based on the available information about the EFT expansion pattern for any
given observable.  To be specific, consider an arbitrary NN
scattering observable $X$ at the center of mass momentum $p$, which is
calculated in chiral EFT up to the order $Q^k$ 
\beq
\label{expansion}
X(p) = X^{(0)} + \Delta X^{(2)} + \Delta X^{(3)} + \ldots + \Delta
X^{(k)} + \Delta
X^{(k+1)} + \ldots  \equiv X^{(k)} + \delta X^{(k)}.
\eeq
The corrections $\Delta X^{(i)}$,  $\Delta X^{(i)} =
\mathcal{O} \big(X^{(0)} Q^i \big)$, are assumed to be known
explicitly up to the order $i =k$. The goal is to estimate the size of neglected
higher-order terms $\delta X^{(k)} = \sum_{n > k} \Delta
X^{(n)}$. We, furthermore, assume that the expansion parameter
$Q$ is given by
\beq
\label{ExpPar}
Q = {\rm max} \bigg( \frac{M_\pi^{\rm eff}}{\Lambda_{\rm b}}, \; \frac{p}{\Lambda_{\rm b}} \bigg)\,.
\eeq
This simple ansatz is motivated by the expectation that at very low
energies, the errors are dominated by the expansion around the
chiral limit. The scale $M_\pi^{\rm eff}$, which will be specified below,
is related to the pion
mass and controls the convergence rate of the expansion around the
chiral limit. At higher energies one would, on the other hand, expect the
expansion to be dominated by powers of momenta. This simple picture
is in qualitative agreement with the error plots for NN phase shifts \cite{Epelbaum:2014efa},
which show clearly the two different regimes mentioned above, see Ref.~\cite{Epelbaum:2015pfa} for a discussion. 

The algorithm proposed by Epelbaum, Krebs and Mei{\ss}ner (EKM) in
Ref.~\cite{Epelbaum:2014efa} employs 
$M_\pi^{\rm eff} = M_\pi$ and 
$\Lambda_{\rm b} = 600$~MeV based on the estimation from the error plots.  
It also assumes 
the truncation error $\delta X^{(k)} $ to be  dominated by the first
neglected term. The truncation errors at orders $Q^i$, $0 \le i \le
k$, are then estimated via 
\begin{equation}
\label{ErrorOrig}
\delta X^{(0)} = Q^2 | X^{(0)}|, \quad \quad \delta X^{(i)} =
\max_{2 \le j \le i} \Big( Q^{i+1} | X^{(0)} |, \; Q^{i+1-j} | \Delta
X^{(j)} | \Big) \; \;  \mbox{for} \; \; i \ge 2\,, 
\end{equation}
subject to the additional constraint 
\begin{equation}
\label{ErrorOrig2}
\delta X^{(i)} \ge \max_{j, m = i, \ldots , k} \Big( | X^{(j)} - X^{(m)} | \Big)\,,
\end{equation}
which ensures that the estimated errors cannot be smaller than the
known actual higher-order contributions.  Notice that this relation
leads, per construction, to overlapping errors at different orders.  
In Ref.~\cite{Binder:2015mbz}, the method was adjusted to make it applicable to
incomplete calculations of few-body observables based on NN
interactions only. 
The EKM approach was applied to a broad range of low-energy reactions
in the single-baryon
\cite{Siemens:2016hdi,Yao:2016vbz,Siemens:2017opr,Blin:2018pmj} as
well as 
few- and many-nucleon 
\cite{Epelbaum:2014sza,Hu:2016nkw,Skibinski:2016dve,Epelbaum:2018ogq,NevoDinur:2018hdo}
sectors.
The robustness of this method and some alternative algorithms 
are discussed in Ref.~\cite{Binder:2018pgl}.
The obvious drawback of the EKM approach is that the
estimated uncertainties do not offer a statistical interpretation. 

In Refs.~\cite{Furnstahl:2015rha,Melendez:2017phj,Melendez:2019izc}, a
more general and statistically well-founded Bayesian approach was developed
to calculate the probability distribution function (pdf) for truncation
errors in chiral EFT. The EKM
approach was then shown to correspond to one particular
choice of prior probability distribution for the coefficients in the
chiral expansion of $X (p)$. In Ref.~\cite{Furnstahl:2015rha}, the EKM
uncertainties for the np total cross section were found to
be consistent with $68\%$  degree-of-belief (DoB) intervals.
The authors of that paper, furthermore, found using the semilocal
coordinate-space regularized (SCS) potentials of Refs.~\cite{Epelbaum:2014efa,Epelbaum:2014sza} the assumed value of
the breakdown scale of $\Lambda_{\rm b} = 600$~MeV to be statistically
consistent for not too soft regulator values, see also Ref.~\cite{Melendez:2017phj} for a
related discussion. Recently, a slightly modified version of the
Bayesian approach developed in
Refs.~\cite{Furnstahl:2015rha,Melendez:2017phj} was applied by the LENPIC
Collaboration to study NN and 3N scattering \cite{Epelbaum:2019zqc}. Below, we briefly outline
the Bayesian model $\bar C_{0.5-10}^{650}$ proposed in that paper,
which will be employed throughout section~\ref{sec6}. For more details
on the Bayesian approach the reader is referred to the original
publications \cite{Furnstahl:2015rha,Melendez:2017phj}.

We begin with rewriting Eq.~(\ref{expansion}) in terms of dimensionless
expansion coefficients $c_i$ by introducing a (generally dimensionfull)
scale $X_{\rm ref}$ via
\beq
\label{coeffci}
X = X_{\rm ref} \left( c_0 + c_2 Q^2 + c_3 Q^3 + c_4 Q^4 + \ldots \right) \,,
\eeq
where\footnote{No meaningful uncertainty estimation
  can be carried out within the Bayesian approach at LO.}   
\beq
\label{tempBE}
X_{\rm ref} = \left\{ \begin{array}{l} X_{\rm ref} = \max \left( | X^{(0)} |, \;  Q^{-2}| \Delta X^{(2)}
                       | \right) \quad \mbox{for} \quad k=2 \, , \\[6pt]
                       X_{\rm ref} = \max \left( | X^{(0)} |, \;  Q^{-2}| \Delta X^{(2)}
    |, \;  Q^{-3}| \Delta X^{(3)}
                       | \right) \quad \mbox{for} \quad k \ge 3\, .
                       \end{array} \right.
\eeq
This choice of the reference scale was found in Ref.~\cite{Epelbaum:2019zqc} to be more robust for
observables that depend on continuously varying parameters, as
compared with the choice of $X_{\rm ref} =  | X^{(0)} |$ adopted in
Ref.~\cite{Melendez:2017phj}. Alternatively, correlations between observables (and thus
the coefficients $c_i$) evaluated at
different values of continuously varying parameters can be taken into
account using Gaussian processes \cite{Melendez:2019izc}. 
Except for the coefficient $c_m =1$, $m \in \{0, 2, 3\}$, used to
set the scale  $X_{\rm ref}$, the expansion coefficients $c_i$ are
assumed to be distributed according to some common pdf ${\rm
  pr} (c_i | \bar c )$  with a hyperparameter $\bar c$.  Performing 
marginalization over $h$
chiral orders $k+1, \ldots, k+h$, which are assumed to dominate the truncation
error,
the probability distribution for the
dimensionless residual $\Delta_k \equiv \sum_{n=k+1}^\infty c_n Q^n
\simeq  \sum_{n=k+1}^{k+h} c_n Q^n$
to take a value $\Delta_k = \Delta$, given the knowledge of $\{c_{i
  \le k} \}$, is given by \cite{Melendez:2017phj}
\begin{equation}
  \label{posteriorGeneral}
{\rm pr}_h ( \Delta  | \{ c_{i \le k} \}) = \frac{\int_0^\infty d \bar c \,
  {\rm pr}_h (\Delta | \bar c ) \,  {\rm pr} (\bar c ) \prod_{i\in A}
  {\rm pr} (c_i | \bar c ) }{\int_0^\infty  d \bar c  \,  {\rm pr} (\bar c )
  \prod_{i\in A}  {\rm pr} (c_i | \bar c ) }\,,
\end{equation}
where the set $A$ is defined as $A = \{n
\in \mathbb{N}_0 \, | \, n \leq k \, \land \, n \neq 1 \, \land \, n \neq m \}$ and
 \begin{equation}
 {\rm pr}_h (\Delta | \bar c ) \equiv \left[ \prod_{i=k+1}^{k+h}
   \int_{-\infty}^\infty d c_i \,  {\rm pr} (c_i | \bar c ) \right]
 \, \delta \bigg( \Delta - \sum_{j=k+1}^{k+h} c_j Q^j \bigg)\,.
 \end{equation}
The model $\bar C_{0.5-10}^{650}$ utilizes 
the 
Gaussian prior of ``set C'' from Ref.~\cite{Melendez:2017phj},
\begin{equation}
  \label{prior}
 {\rm pr} (c_i | \bar c ) = \frac{1}{\sqrt{2 \pi} \bar c} \, e^{-
   c_i^2/(2 \bar c^2 )}, \quad \quad {\rm pr} ( \bar c ) = \frac{1}{\ln ( \bar c_> / \bar c_< )} \, 
\frac{1}{\bar c} \, \theta (\bar c - \bar c_< ) \, \theta (\bar c_> -
\bar c )\,,
\end{equation}  
for which the integrals in Eq.~(\ref{posteriorGeneral}) can be
performed analytically \cite{Melendez:2017phj}, and uses the values of $h = 10$,
$\bar c_< = 0.5$ and $\bar c_> = 10$. Following Ref.~\cite{CD18-EE},
the scales that control the expansion parameter are set to
$M_\pi^{\rm eff} = 200$~MeV and  $\Lambda_{\rm b} = 650$~MeV. 
The sensitivity of the estimated uncertainties to the choice of prior pdf is discussed in
Refs.~\cite{Furnstahl:2015rha,Melendez:2017phj,Epelbaum:2019zqc}. One
generally finds minor dependence on the prior pdf if a sufficient
amount of information on the coefficients $c_i$ is available.

\section{Selected results}
\label{sec6}

\subsection{The two-nucleon system}
\label{NNresults}

We now turn to the calculation of phase shifts and observables in the
two-nucleon system. While the derivation and regularization of the
nuclear forces have been outlined in the previous sections, we also
need to specify the numerical values of all relevant physical quantities and
LECs
to perform actual calculations. For 
pion-exchange contributions to the potential, all LECs can be extracted
from processes involving at most one nucleon, making it parameter-free.
In the TPEP, we use the values of the $\pi$N
LECs as determined recently by matching the $\pi$N scattering
amplitude from chiral perturbation theory to a Roy-Steiner equations
analysis of $\pi$N scattering data at the subthreshold point
\cite{Hoferichter:2015tha}. 

We account for the
isospin-breaking effects due to the different pion masses in the OPEP
and employ the physical masses of the charged and neutral pions
$M_{\pi^\pm} = 139.57$ MeV and $M_{\pi^0} = 134.98$ MeV,  while
we use the isospin-averaged value of $M_\pi = 138.03$ MeV in the
TPEP. We adopt an effective value of $g_A = 1.29$ for the nucleon
axial coupling constant which is slightly larger than the current
experimental average value of $g_A = 1.2723(23)$
\cite{Tanabashi:2018oca} because it already accounts for the
Goldberger-Treiman discrepancy, see Ref.~\cite{Fettes:1998ud} for a
related discussion. The employed value of the pion decay constant is
$F_\pi = 92.4$ MeV.

In contrast to the parameter-free long-range potential, the
short-range contact interactions in the two-nucleon force have to be
determined from experimental  NN data. In order to achieve a proper
reproduction of pp and, to a lesser extent, np
scattering data, it is crucial to also include electromagnetic
interactions between the nucleons. Although these interactions are
accompanied by powers of a numerically small coupling constant $\alpha
\sim 1/137$, they are enhanced at low energies and/or forward angles
due to the infrared singularity of the photon propagator or,
equivalently, due to their long-range nature. Here, we follow the
treatment of the Nijmegen group \cite{Stoks:1993tb}
and include the so-called improved Coulomb potential
\cite{Austen:1983te}, the magnetic-moment interaction
\cite{Stoks:1990us} as well as the vacuum-polarization potential
\cite{Durand:1957zz} in the calculation of proton-proton
observables. The magnetic moment interaction is also taken into
account in neutron-proton scattering. To the best of our knowledge,
these effects have been included in every partial-wave analysis (PWA) of or
fit of a high-quality potential model from NN data since the
Nijmegen PWA of Ref.~\cite{Stoks:1993tb}, so that differences in their predictions stem from
modeling the strong interaction, the experimental input and/or
details of the fitting procedure itself. 

For scattering data we use the Granada-2013 database
\cite{Perez:2013jpa} which consists of experimental data for NN
elastic scattering up to $E_{\rm lab} = 350$ MeV from 1950 up to
2013.\footnote{Strictly speaking, our database differs from the one of
  Ref.~\cite{Perez:2013jpa} by the omission of the data set from
  Ref.~\cite{Daub:2012qb}, see Ref.~\cite{Reinert:2017usi} for more
  details.}
The database contains the data that have
been found to be mutually compatible by means of a $3\sigma$ rejection
criterion in the corresponding phase shift analysis of
Ref.~\cite{Perez:2013jpa}.  
The presence of very precisely measured proton-proton data in the
database, such as those of Ref.~\cite{Cox:1968jxz}, motivated us to
introduce the N$^4$LO$^+$ version of the potential. As the proper
description of these data requires a precise reproduction of F-waves,
the N$^4$LO$^+$ potential adds the four leading F-wave contact
interactions 
\begin{equation}
	\langle ^SF_j, \, p' | V_{\rm cont} | ^SF_j, \, p \rangle  =  E_{SFj} \, p^3 p'^3 \,,
\end{equation}
formally appearing at N$^5$LO and entering the $^3F_2$, $^1F_3$,
$^3F_3$ and $^3F_4$ partial waves, to the N$^4$LO potential.

The fits have been performed for all cutoffs $\Lambda = 400, 450, 500$
and $550$ MeV as well as for all orders from LO up to
N$^4$LO$^+$.\footnote{In our paper \cite{Reinert:2017usi} also the cutoff  $\Lambda =
  350$~MeV was considered. Given the sizable finite-$\Lambda$
  artifacts for this very soft cutoff choice, we do not consider this
  case in the following discussion.} When determining the values of the contact LECs, one has
to decide up to which energy $E_{\rm lab}$ the experimental data
should be taken into account. One is faced with the two competing 
features: on the one hand, the inclusion of as many data as possible is desirable
from a data fitting point of view. On the other hand, the truncation errors for the
chiral interactions become larger at high energies. We therefore chose
the maximum energy $E_{\rm lab}$ of data to be included to be $E_{\rm
  max} = 260$ MeV for N$^4$LO and N$^4$LO$^+$, while we reduced the
energy to $E_{\rm max} = 25, 100, 125$ and $200$ MeV at the orders LO,
NLO, N$^2$LO and N$^3$LO, respectively. From N$^3$LO on, we also
adjust the deuteron binding energy $B_d$ and the coherent
neutron-proton scattering length $b_{\rm np}$ to reproduce their
experimental values of $B_d = 2.224575(9)$ MeV
\cite{VanDerLeun:1982bhg} and $b_{\rm np} = -3.7405(9)$ fm
\cite{Schoen:2003my}. 

Table \ref{tab:chi2_AllOrders} shows the reproduction of
neutron-proton and proton-proton scattering data in terms of
$\chi^2$/datum values at all considered orders for the cutoff
$\Lambda = 450$ MeV.\footnote{We have corrected the last figures in
  the values for $\chi^2/{\rm datum}$ for np data in the
  $E_{\rm lab}$ bins of 0--100~MeV and 0--200~MeV at N$^3$LO and
  N$^4$LO$^+$ quoted in Table 3 of Ref.~\cite{Reinert:2017usi}.} 
\begin{table}[t]
    \caption{$\chi^2$/datum for the description of the neutron-proton and proton-proton scattering data at various orders
        in the chiral expansion for $\Lambda = 450$~MeV. The numbers in brackets after the order indicate the number of parameters entering 
        the neutron-proton and proton-proton potentials.
        \label{tab:chi2_AllOrders}}
    \smallskip
    \begin{tabular*}{\textwidth}{@{\extracolsep{\fill}}lllllll}
        \hline 
        \hline 
        \noalign{\smallskip}
        $E_{\rm lab}$ bin &  LO$_{(3)}$   &  NLO$_{(10)}$   &  N$^2$LO$_{(10)}$   &  N$^3$LO$_{(22)}$  &  N$^4$LO$_{(23)}$  & N$^4$LO$^+$$_{(27)}$  
        \smallskip
        \\
        \hline 
        \hline 
        \multicolumn{5}{l}{neutron-proton scattering data} \\ 
        0--100 & 73 & 2.2 & 1.2 & 1.07 & 1.07 & 1.07\\ 
        0--200 & 62 & 5.4 & 1.7 & 1.09 & 1.08 & 1.06\\
        0--300 & 75 & 14  & 4.2 & 2.01 & 1.16 & 1.06\\ [4pt]
        \hline 
        \multicolumn{5}{l}{proton-proton scattering data} \\ 
        0--100 & 2290 & 10 & 2.2 & 0.90 & 0.88 & 0.86\\ 
        0--200 & 1770 & 90 & 37  & 1.99 & 1.42 & 0.95\\ 
        0--300 & 1380 & 90 & 41  & 3.43 & 1.67 & 1.00\\ [4pt]
        \hline 
        \hline 
    \end{tabular*}
\end{table}
As expected, a clear order-by-order improvement in the description of
the scattering data can be seen. Table \ref{tab:chi2_AllOrders} also
gives the number of adjustable parameters at each order which also
includes isospin-breaking LECs contributing to the $^1S_0$ partial
wave. It should be noted that no new contact interactions are added
when going from NLO to N$^2$LO and that the observed improvement of
the $\chi^2$/datum values is entirely due to the N$^2$LO contributions
to the parameter-free TPEP. A similar situation occurs when going from
N$^3$LO to N$^4$LO, although here we also allow for additional
isospin-breaking of the $C_{1S0}$ contact LEC splitting it into two
independently adjustable parameters for the neutron-proton and
proton-proton/neutron-neutron systems. These improvements demonstrate
both the importance of the chiral TPEP in the nuclear force and
the predictive power of
chiral perturbation
theory, which allows to use LECs extracted in one process for making
parameter-free predictions for (parts of) another. 

Starting from N$^3$LO, a satisfactory description of the
neutron-proton data in the energy range of $E_{\rm lab} = 0-200$ MeV
and the proton-proton data for $E_{\rm lab} = 0-100$ MeV is
achieved. Although the N$^4$LO potential improves on this, especially
at intermediate and higher energies, it does not achieve a
$\chi^2$/datum~$\sim$~$1$ description of the proton-proton data for
$E_{\rm lab} \ge 100$~MeV. In the intermediate region,
this value is significantly affected by the already
mentioned high-precision data which requires an accurate description
of F-waves. At N$^4$LO the differential cross section data set of
Ref.~\cite{Cox:1968jxz} at $E_{\rm lab} = 144.1$~MeV, although well
described within the Bayesian truncation errors, yields a
$\chi^2$/datum value of $27.88$.  

The situation is much improved once we switch to the N$^4$LO$^+$
potential and short-range interactions in F-waves are added. The
description of scattering data at higher energies is generally
improved and also the high-precision proton-proton data at
intermediate energies is accurately reproduced leading to a
$\chi^2$/datum~$\sim$~1 description of the complete scattering
database. Throughout the orders LO $-$ N$^4$LO the $\chi^2$/datum
value for proton-proton scattering up to 200 or 300 MeV has been
larger than the one for neutron-proton scattering. This is plausible
as proton-proton data is in general more precise than neutron-proton
data and because only isovector partial waves contribute to it and
hence only roughly half of the total number of parameters. However, at
N$^4$LO$^+$, the reproduction of proton-proton data becomes very
accurate while the slightly larger $\chi^2$/datum values for the
neutron-proton data as compared to proton-proton data
reflects the larger statistical fluctuations among
different data sets. This can be seen as an indication for reaching
the threshold where the model accuracy approaches the precision of the
data. In fact, the description of the scattering data at N$^4$LO$^+$
and $\Lambda = 450$~MeV is comparable to or exceeds that of the
high-quality semi-phenomenological potentials such as CD-Bonn
\cite{Machleidt:2000ge}, Nijm I, II \cite{Stoks:1994wp} and Reid93
\cite{Stoks:1994wp}. Thanks to the parameter-free effects of the TPEP
this is achieved with only 27 adjustable short-range parameters
instead of the $\sim$~40~$-$~50 parameters used in those potentials.  

\begin{figure}[tb]
    \begin{center}
        \includegraphics[width=1.0\textwidth,keepaspectratio,angle=0,clip]{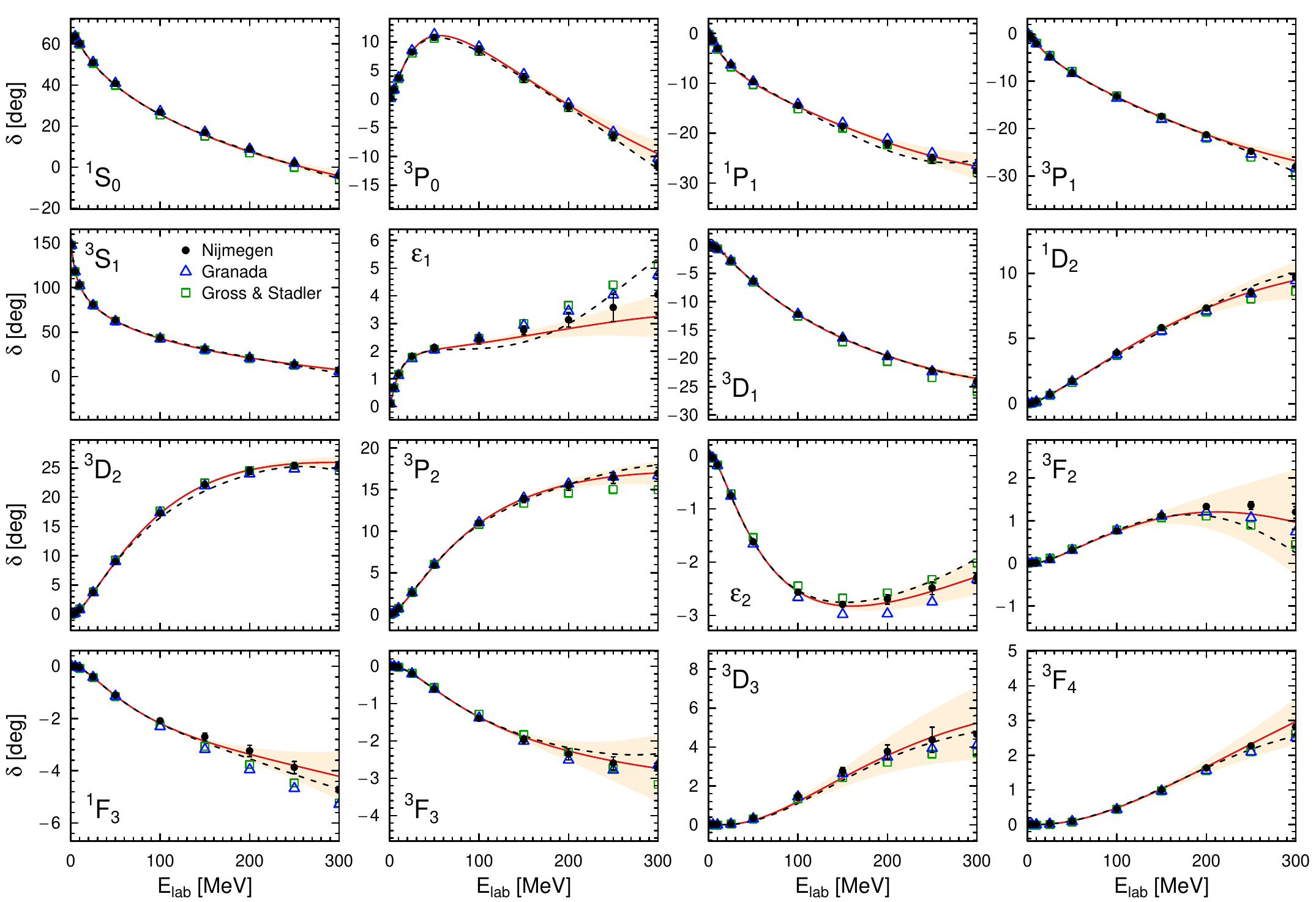}
    \end{center}
    \caption{
        Neutron-proton phase shifts with respect to Riccati-Bessel
        functions in comparison with the Nijmegen \cite{Stoks:1993tb}
        (solid dots), the Granada \cite{Perez:2013mwa} (blue open
        triangles) and Gross-Stadler \cite{Gross:2008ps} (green open
        squares) PWA. Red solid lines and peach-colored bands denote
        the 
        central results and 68\% DoB truncation errors at the order
        N$^4$LO$^+$ for the cutoff $\Lambda = 450$~MeV. Black
        dashed lines denote the result of the nonlocal N$^4$LO$^+$ potential of
        Ref.~\cite{Entem:2017gor} for the cutoff $\Lambda
        = 500$~MeV.    The shown uncertainties of the Nijmegen PWA
        correspond to systematic errors defined in Eq.~(32)
       of Ref.~\cite{Epelbaum:2014efa}.  
        \label{fig:phaseshifts-450MeV-np}
    }
\end{figure}

\begin{figure}[tb]
    \begin{center}
        \includegraphics[width=1.0\textwidth,keepaspectratio,angle=0,clip]{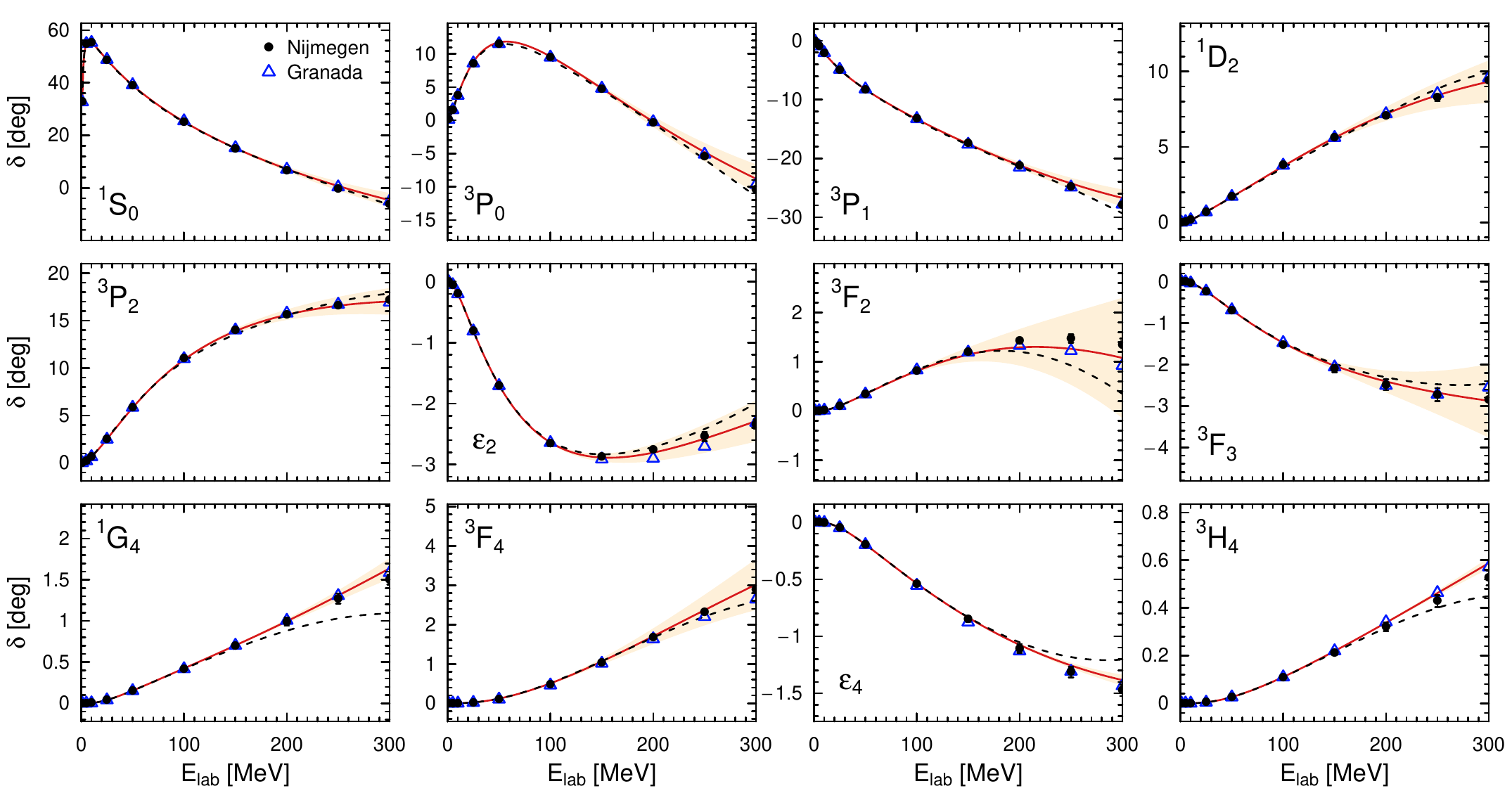}
    \end{center}
    \caption{Proton-proton phase shifts with respect to Coulomb wave
      functions in comparison with the Nijmegen \cite{Stoks:1993tb}
        (solid dots) and the Granada \cite{Perez:2013mwa} (blue open
        triangles) PWA. Red solid lines and peach-colored bands denote the
      central result and 68\% DoB truncation errors at the order
      N$^4$LO$^+$ for the cutoff $\Lambda = 450$~MeV. Black dashed
      lines denote the result of the nonlocal N$^4$LO$^+$ potential of
      Ref.~\cite{Entem:2017gor} for the cutoff $\Lambda =
      500$~MeV.    The shown uncertainties of the Nijmegen PWA
        correspond to systematic errors defined in Eq.~(32)
       of Ref.~\cite{Epelbaum:2014efa}.  
        \label{fig:phaseshifts-450MeV-pp}
    }
\end{figure}

Indeed, due to the excellent description of the data, the obtained results
at $\Lambda = 450$~MeV qualify to be considered a partial-wave
analysis. In Figs.~\ref{fig:phaseshifts-450MeV-np} and
\ref{fig:phaseshifts-450MeV-pp} we show the obtained N$^4$LO$^+$
neutron-proton and proton-proton phase shifts for $\Lambda = 450$~MeV,
respectively. We compare them to the 2013 Granada analysis
\cite{Perez:2013jpa} and in the case of neutron-proton scattering also
to the corresponding 2008 analysis by Gross and Stadler
\cite{Gross:2008ps}. Furthermore, we also show the predictions from
the N$^4$LO$^+$ potential of Ref.~\cite{Entem:2017gor} at the
intermediate cutoff $\Lambda = 500$~MeV. 

In general, there is good agreement between the shown N$^4$LO$^+$
phase shifts and the results obtained by the considered phase shift
analyses. This is especially true for the case of proton-proton
phase shifts which are more strongly constrained by the precise
experimental data. Some discrepancies among the different results
remain e.g. around the maximum of the ${}^3P_0$ phase shift where the
N$^4$LO$^+$ prediction for the proton-proton phase is slightly larger
than the ones of the Nijmegen and Granada PWAs, resulting in a $\sim
3\sigma$ deviation from the former at $E_{\rm lab} = 50$~MeV. On the
other hand, our neutron-proton phase shifts fall in between the
results of the two PWAs.
The
study of isospin-breaking effects in P-waves beyond the ones included
in the two PWAs and the current version of the semilocal
momentum-space regularized (SMS) interaction of Ref.~\cite{Reinert:2017usi} is
expected to shed some light on this issue. We can also compare our
results at N$^4$LO$^+$ to the ones of
Ref.~\cite{Entem:2017gor}. Similar to the comparison with the PWAs,
agreement with proton-proton phases is better than with neutron-proton
ones. There are, however, notable differences in the ${}^3P_0$,
${}^3P_2$ and ${}^3D_2$ waves starting at low or intermediate
energies. At higher energies around $E_{\rm lab} = 250-300$~MeV, 
a change in curvature of the phase shift as a function of energy is
visible e.g. in the ${}^1P_1$ and ${}^3P_1$ waves, which is presumably
caused by the regulator employed in Ref.~\cite{Entem:2017gor}. The effects of regulator artifacts
can be observed particularly well in the ${}^1G_4$, ${}^3H_4$ and
$\epsilon_4$ phase shifts and mixing angle shown in
Fig.~\ref{fig:phaseshifts-450MeV-pp} since they do not involve any
adjustable short-range parameters at N$^4$LO$^+$ but are solely
determined by the long-range pion-exchange potential. Here, the local
regulator of Eq.~\eqref{eq:pion-propagator-form-factor} leads to an
undistorted reproduction of the peripheral phase shifts. 

\begin{figure}[tb]
    \begin{center}
        \includegraphics[width=\textwidth,keepaspectratio,angle=0,clip]{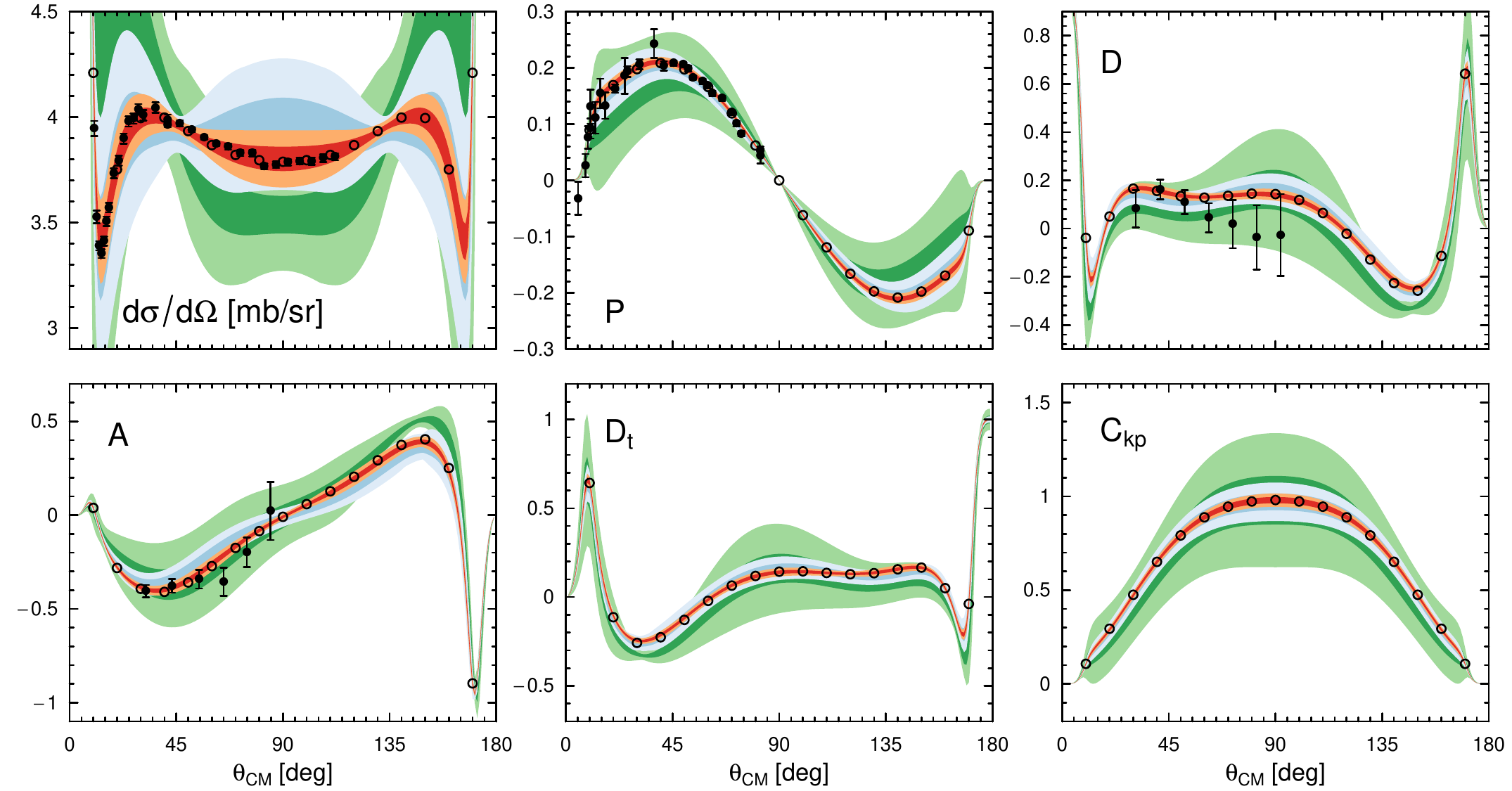}
    \end{center}
    \caption{
    	Selected proton-proton observables around $E_{\rm lab} =
        143$~MeV: Differential cross section $d\sigma/d\Omega$ at
        $E_{\rm lab} = 144.1$~MeV with experimental data taken from
        Ref.~\cite{Cox:1968jxz} and Ref.~\cite{Jarvis:1971fla}. The
        data sets have been corrected for their estimated norms of
        $0.988$ and $1.001$, respectively. Analyzing power $P$ at
        $E_{\rm lab} = 142$~MeV with experimental data taken from
        Ref.~\cite{Taylor:1960}. The data have been floated and multiplied
        by an estimated norm of $0.942$. Depolarization $D$, rotation
        parameter $A$,
        polarization transfer coefficient $D_t$ and spin-correlation
        parameter $C_{kp}$ at $E_{\rm lab} = 143$~MeV with
        experimental data taken from Refs.~\cite{Bird:1961} and
        \cite{Jarvis:1963}. The light- (dark-) shaded  green,
        blue and red bands depict the 68\% (95\%) DoB truncation
        errors at N$^2$LO, N$^3$LO and N$^4$LO$^+$,
        respectively. Open circles show the predictions of the
        Nijmegen partial-wave analysis \cite{Stoks:1993tb}. 
        \label{fig:observables-pp}
    }
\end{figure}

Selected proton-proton scattering observables and their estimated
truncation error at various orders are shown in
Fig.~\ref{fig:observables-pp} for $E_{\rm lab}$ around $\sim
143$~MeV. In particular, we show our predictions for the differential
cross section at $E_{\rm lab} = 144.1$~MeV and compare them with two
high-precision data sets, most notably the one of
Ref.~\cite{Cox:1968jxz}, which motivated the introduction of the
N$^4$LO$^+$ potential as discussed above. The data are well described
within the given truncation error for all considered orders, but the
N$^4$LO$^+$ clearly allows for a proper quantitative
description. Likewise, the reproduction of the spin observables in
Fig.~\ref{fig:observables-pp} is excellent already at N$^3$LO with a
good convergence pattern. Notice however, that the error bands at
lower orders for $D$ ($A$) at the minimum (maximum) around
$\Theta_{\rm CM} = 150^\circ$ do not overlap with the ones for 
N$^{\ge 3}$LO and are indeed underestimating the uncertainty. Here we find
that the value of the observable in that particular angular region is
notably shifted starting at N$^3$LO while lower-order corrections are small,
such that the overall scale in Eq.~(\ref{tempBE}) is still
underestimated. Using a more sophisticated Bayesian
approach of Ref.~\cite{Melendez:2019izc} would likely allow for a more reliable
estimation of the truncation errors at LO-N$^2$LO in these particular
cases.  

\begin{figure}[tb]
    \begin{center}
        \includegraphics[width=1.0\textwidth,keepaspectratio,angle=0,clip]{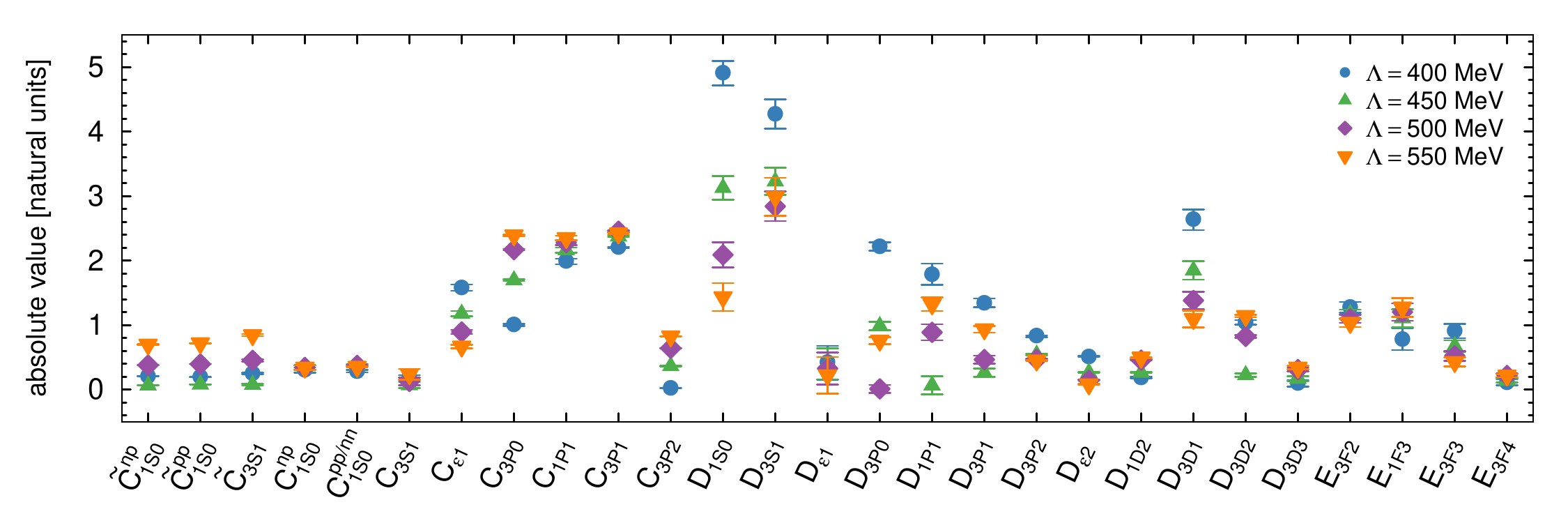}
    \end{center}
    \caption{  
    	Absolute values of the contact interaction LECs in natural
        units at the order N$^4$LO$^+$ for all considered
        cutoffs. Error bars represent the statistical errors of the
        LECs. 
        \label{fig:NaturalLECs}
    }
  \end{figure}

There are various a posteriori checks that can be performed to test
the self-consistency and quality of the fit. First, the values of the
LECs have to be of natural size assuming the cutoff is kept below the
hard scale. The expected sizes of the spectroscopic contact LECs can
be estimated to be \cite{Epelbaum:2014efa} 
\begin{equation}
	\label{eq:natural-sizes}
    | \tilde C_i | \sim \frac{4 \pi}{F_\pi^2}, \qquad
    | C_i | \sim \frac{4 \pi}{F_\pi^2 \Lambda_{\rm b}^2}, \qquad
    | D_i | \sim \frac{4 \pi}{F_\pi^2 \Lambda_{\rm b}^4}, \qquad
    | E_i | \sim \frac{4 \pi}{F_\pi^2 \Lambda_{\rm b}^6},
\end{equation}
where the LECs $\tilde{C}_i$, $C_i$, $D_i$ and $E_i$ start to
contribute at order $Q^0$, $Q^2$, $Q^4$ and $Q^6$,
respectively. $\Lambda_{\rm b}$ is the breakdown scale of the chiral
expansion discussed in Sec.~\ref{sec5}. Furthermore,
the factor of $4\pi$ emerges from the angular integration of the
partial-wave decomposition and has been included in the definition of
the spectroscopic LECs. If we now divide the contact LECs obtained in
the fit by their expected sizes in Eq.~\eqref{eq:natural-sizes}, we
consequently should obtain values of unit
magnitude. Fig.~\ref{fig:NaturalLECs} shows the absolute values of the
LECs at N$^4$LO$^+$ in these natural units for all considered values
of the cutoff $\Lambda$ using $\Lambda_{\rm b} = 650$~MeV.  
As can be seen, all LECs are indeed of natural size with $D_{1S0}$ and
$D_{3S1}$ being among the largest in magnitude. This is especially
true for the softest cutoff $\Lambda = 400$ MeV, for which also most of the
other-$Q^4$ LECs turn out to be slightly larger than at higher values
of the cutoff. This indicates that at $\Lambda = 400$ MeV and below,
finite-cutoff artifacts start to increase, leading to a lower
effective breakdown scale compared to the other considered
cutoffs. Notice further that the values for the $Q^6$ LECs $E_i$
included at N$^4$LO$^+$ turn out to be of a perfectly natural size. Therefore,
even though we have emphasized their importance in describing some
high-precision proton-proton data and achieving a
$\chi^2$/datum~$\sim$~1 description of the database, their
actual contributions agree with the expectations from naive dimensional
analysis (i.e.~Weinberg) power counting, and there is no need to
promote them to a lower order. 

In addition to the absolute of the central values,
Fig.~\ref{fig:NaturalLECs} also shows the statistical uncertainties of
the
contact LECs as determined from the covariance matrix of
the fit (expressed in their natural units).
When going from $\tilde{C}_i$, $C_i$, $D_i$ to $E_i$ the
statistical relative errors tend to increase. This is in accordance
with the decreasing importance of higher-order contributions as
predicted by power counting. One also notices that errors are smaller
for LECs entering isovector partial waves, because these parameters are
mainly constrained by the more precise proton-proton data. Since we
perform a combined fit of neutron-proton and proton-proton data, the
isovector partial waves are not only constrained by more precise data
but also by more data in general compared to the isoscalar partial
waves which have to be extracted from neutron-proton data alone. The
covariance matrix also gives access to the correlations among the
LECs. As to be expected, correlations mostly occur among LECs entering
the same partial waves with the largest ones arising in the channels
with the most parameters, namely in the $^1S_0$ and $^3S_1 - {}^3D_1$
channels. Nevertheless, all LECs are well-constrained as can already
be seen by looking at the errors in Fig.~\ref{fig:NaturalLECs}. We can
further look at the largest eigenvalue of the covariance matrix of the
natural LECs as a measure of how well-determined the parameters
are. Throughout the considered range of the cutoff $\Lambda = 400
- 550$~MeV, the largest eigenvalue of the covariance matrix does not
exceed $0.1$ and is $\sim 0.08$ for $\Lambda = 450$~MeV. 

From the point of view of data fitting, another check concerns the
statistical assumptions underlying a $\chi^2$ fit. One usually assumes
that the residuals $r_i = (O_i^{\rm exp}-O_i^{\rm th})/\Delta O_i$
follow a normal distribution $\mathcal{N}(0,1)$ with zero mean and
unit standard deviation. Here $O_i^{\rm exp}$ and $\Delta O_i$ are the
experimental value and its error of an observable and $O_i^{\rm th}$
is its calculated "theoretical" value. If the assumptions on the
normally-distributed residuals can be verified, this confirms that the
data are described sufficiently well by the theoretical model. An easy
and often employed check is the value of $\chi^2$ per degree of
freedom. For the N$^4$LO$^+$ fit with $\Lambda = 450$ MeV we get
$\chi^2 = 4708.65$ in the fitting range of $E_{\rm lab} = 0 - 260$ MeV
with the number of data $N_{\rm dat} = 4616$ and the number of
parameters $N_{\rm par} = 27$. Consequently, we obtain $\chi^2/\nu =
1.026$ with $\nu = N_{\rm dat} - N_{\rm par}$. If the residuals are
indeed normal-distributed then $\chi^2/\nu$ should follow the
$\chi^2$-distribution and yields $\chi^2/\nu = 1 \pm \sqrt{2/\nu} = 1
\pm 0.021$ as the 68\% confidence interval.

\begin{figure}[tb]
    \begin{center}
        \includegraphics[width=0.6\textwidth,keepaspectratio,angle=0,clip]{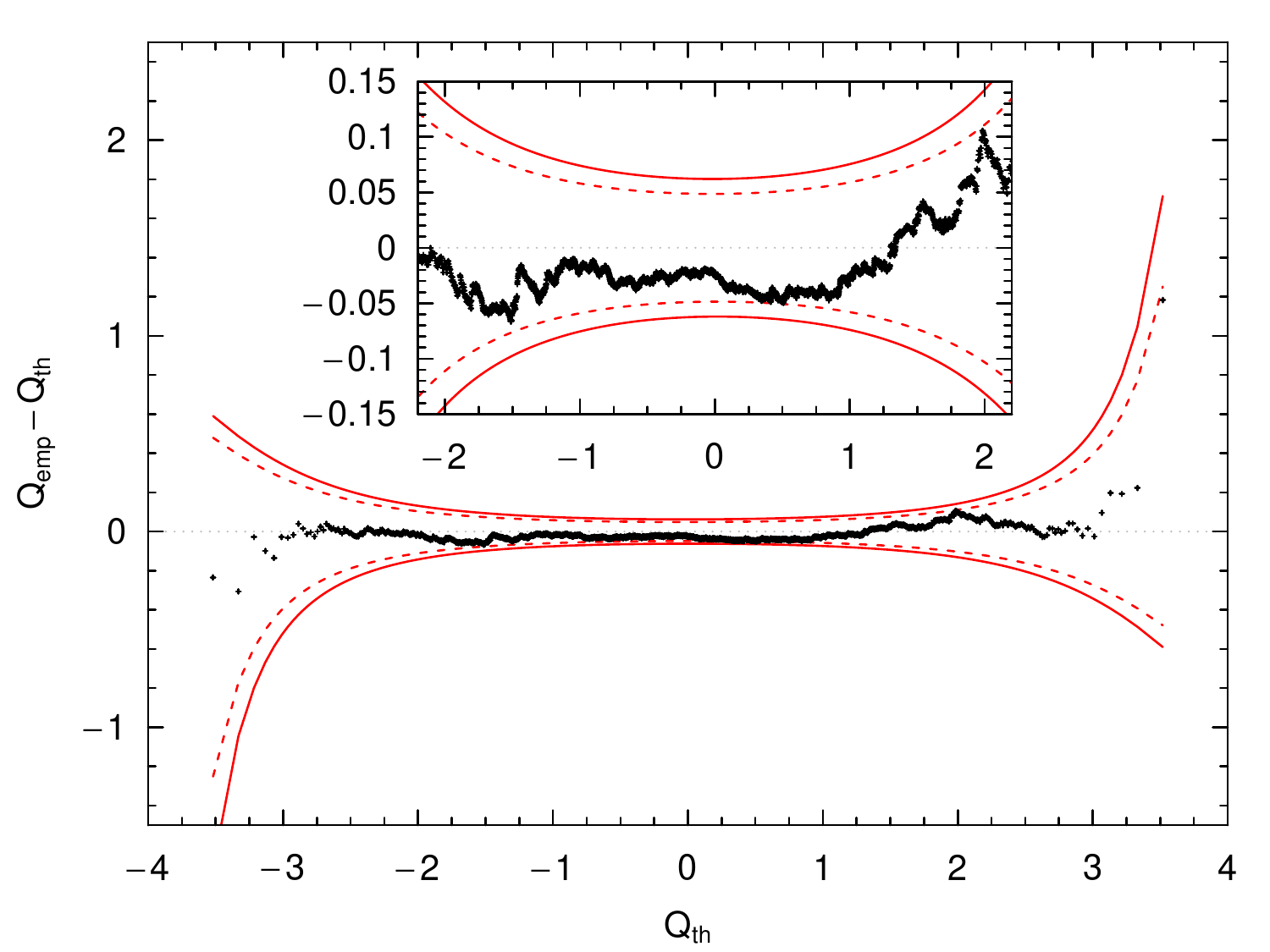}
    \end{center}
    \caption{  
    	Rotated Quantile-Quantile plot of the empirical quantiles at
        N$^4$LO$^+$ and $\Lambda = 450$~MeV versus the quantiles of
        the normal distribution $\mathcal{N}(0,1)$ Dotted (solid) red
        bands denote the 68\% (95\%) confidence bands of the
        tail-sensitive test by Aldor-Noiman et
        al.~\cite{Aldor-Noiman:2013}. 
        \label{fig:q-q-plot}
    }
\end{figure}

We can go one step beyond this simple check and plot the quantiles of
the empirical distribution of residuals $r_i$ that we obtain against
the quantiles of the assumed normal distribution
$\mathcal{N}(0,1)$. If they are the same, they should lie on the
diagonal line $x = y$. In order to statistically quantify deviations
from the diagonal, confidence bands have been derived with one of the
most recent and most sensitive being the ones by Aldor-Noiman et
al.~\cite{Aldor-Noiman:2013}. This graphical test for
normal-distributed residuals has been first applied to the analysis of
nucleon-nucleon scattering by Navarro P\'{e}rez et al.~\cite{Perez:2014yla} and
named the "tail-sensitive test" in that
publication. Fig.~\ref{fig:q-q-plot} shows a rotated quantile-quantile
plot for the N$^4$LO$^+$ residuals at $\Lambda = 450$ MeV where the
theoretical quantiles have been subtracted from the empirical ones on
the $y$-axis, turning the diagonal line into a horizontal one. 
As evident from the figure, the empirical distribution of residuals
lies within the 68\% confidence region of the tail-sensitive test
signaling that the residuals are indeed normal-distributed. The
quantile-quantile plot for the other values of the cutoff turn out to
be overall similar, but perform slightly worse. For $\Lambda =
500$~MeV and $\Lambda = 550$~MeV the quantiles that are already close
to the edge of the 68\% confidence region in Fig.~\ref{fig:q-q-plot}
cross these limits but still stay well within the 95\% confidence
region. The increased cutoff-artifacts at $\Lambda = 400$~MeV manifest
themselves in a stronger deviation from normality as the plotted
quantiles also cross the 95\% confidence limits at the spike at
$Q_{\rm th} = 2$ in Fig.~\ref{fig:q-q-plot}. 

We now turn to the extended error analysis for observable predictions. While the truncation of the chiral expansion is clearly the dominant source of uncertainty at higher energies, other sources of uncertainty can become relevant at N$^4$LO$^+$. In particular we account for the following sources of uncertainty:
\begin{itemize}
\item{
    \emph{Statistical uncertainties of NN LECs:} As already mentioned,
    Fig.~\ref{fig:NaturalLECs} shows the statistical errors of the
    contact LECs as determined from the covariance matrix of the
    fit. The uncertainties of the parameters can then be propagated
    from the covariance matrix to the observable of interest. While it
    is always possible to do this via a Monte Carlo sampling of the
    corresponding multivariate Gaussian probability distribution, it
    is computationally much more convenient to do a Taylor expansion
    of the desired observable with respect to the LECs and evaluate
    the moments of the LECs analytically. While a linear expansion is
    commonly employed, it has been argued in
    Ref.~\cite{Carlsson:2015vda} that some observables require a
    second order expansion for an accurate reproduction of their
    uncertainties. In the case of large second-order contributions,
    the error bars become asymmetric and we usually give both the
    upper and lower error to accommodate for this possibility. 
    Notice that in such a case, the probability density of the observable
    is not Gaussian and the quoted uncertainties do not necessarily
    correspond to a $68$\% degree-of-belief. 
} 
\item{
    \emph{Statistical uncertainties of $\pi$N LECs:} In addition to
    the central values, the authors of Ref.~\cite{Hoferichter:2015tha}
    also give the covariance matrix as determined from $\pi$N
    scattering data. Propagation of these uncertainties to NN
    observables is, however, less straightforward, because the values
    of the NN contact interactions depend on the values of $\pi$N
    LECs. We thus resort to some Monte Carlo sampling of the
    multivariate Gaussian probability distribution of the $\pi$N LECs
    given by their central values and their covariance matrix. For
    each of the sampled sets of LECs, we refit the NN contact LECs
    before calculating any observables. The uncertainty of a given 
    observable can then be estimated in a standard way from the
    variance of the results calculated with different $\pi$N LEC
    sets. Due to the need to refit the contact interactions for each
    sampled set of $\pi$N LECs and the computational overhead related
    to it, we have restricted the total number of such sets to
    50. Although this is a quite low statistics for a Monte Carlo
    approach, it should give an idea of the order of magnitude of the
    uncertainty. It indeed turns out that the uncertainty related to
    the statistical error of the $\pi$N LECs is small compared to the
    other sources of uncertainty. However, the aforementioned approach
    does not probe the systematic errors in the determination of the
    $\pi$N LECs emerging from the truncation of the chiral expansion and thus does not
    represent the full uncertainty related to these LECs.  
}
\item{
    \emph{Uncertainty due to the choice of the maximum fit energy:} The extracted values
    of the contact LECs also depend on details of the fitting
    protocol. In particular, we probe the impact of the choice for the
    maximum laboratory energy $E_{\rm max} = 260$~MeV up to which
    scattering data is included in the N$^4$LO$^+$ fit. This is
    achieved by performing additional fits with $E_{\rm max} =
    220$~MeV and $E_{\rm max} = 300$~MeV and determining the maximum
    deviation of the observables from the $E_{\rm max} = 260$~MeV
    predictions. Unlike the aforementioned uncertainties, the error
    estimated via this simple procedure does not reflect any
    particular degree-of-belief. 
}
\end{itemize}

\begin{figure}[tb]
    \begin{center}
        \includegraphics[width=\textwidth,keepaspectratio,angle=0,clip]{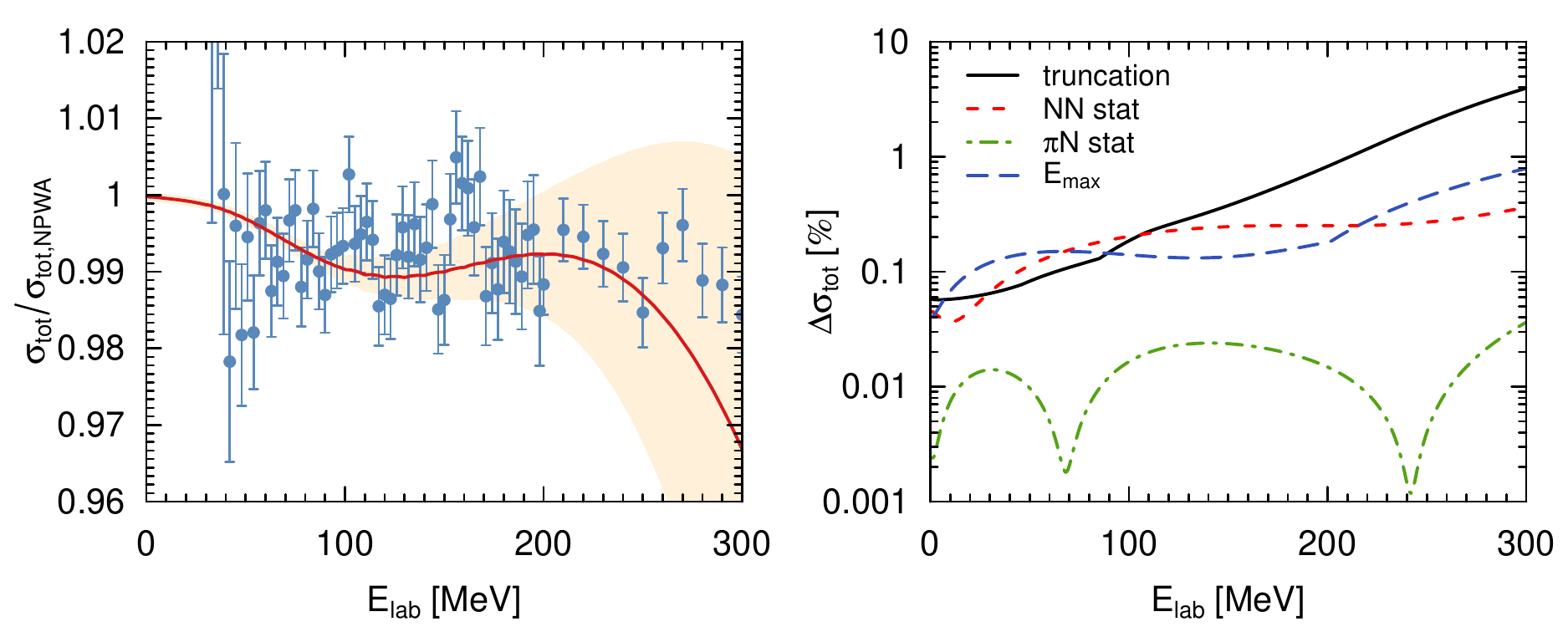}
    \end{center}
    \caption{
        \label{fig:SGT-errors}
        Neutron-proton total cross section in the range of $E_{\rm
          lab}=0-300$~MeV. The plot on the left shows the results
        divided by the predictions of the Nijmegen PWA. The red line
        and peach-colored band show the central values and truncation
        errors (the $68$\% DoB interval) at the order N$^4$LO$^+$ and for $\Lambda =
        450$~MeV. The experimental data are taken from
        Ref.~\cite{Lisowski:1982rm} and have been corrected for their
        estimated norm of $0.999$. The plot on the right shows the
        relative uncertainties as discussed in the text. 
    }
\end{figure}

As an example, Fig.~\ref{fig:SGT-errors} shows the neutron-proton
total cross section and the corresponding uncertainties in the energy
range $E_{\rm lab}=0-300$~MeV. 
The plot on the left in Fig.~\ref{fig:SGT-errors} shows the ratio
of our predictions using the N$^4$LO$^+$ potential at $\Lambda =
450$~MeV and the result of the Nijmegen partial-wave analysis
\cite{Stoks:1993tb}. In the right panel, the different relative errors stemming
from the various sources discussed above are shown. For the case of the
statistical errors of the NN contact interactions, second order
effects and resulting asymmetries in the error bands turn out to be
small for the total cross section, and the plotted uncertainty
corresponds to the average of upper and lower statistical errors. As
expected, the dominant contribution to the uncertainty at
higher energies ($E_{\rm lab} > 100$~MeV) arises from the truncation
of the chiral expansion. At lower energies, however, other sources of
uncertainty become relevant and indeed both the statistical errors of
the NN contact LECs and the uncertainty due to the maximum fitting
energy are larger than the truncation error in the range of $E_{\rm
  lab} = 30-100$~MeV. When quantitatively comparing the different
errors, one has to keep in mind that the uncertainty due to the
maximum fitting energy does not correspond to a particular
degree-of-belief. The uncertainty arising from the statistical errors
of the $\pi$N LECs is found to be significantly smaller throughout the
whole considered energy range and is negligible for the total cross
section. Finally, we would like to comment on the origin of the
existing kinks in the right-hand-side plot of
Fig.~\ref{fig:SGT-errors}. In particular, the kink in the $E_{\rm
  max}$-error at around 200 MeV arises because of the maximum
operation. Below 200 MeV, the error is dominated by the deviation of
the $E_{\rm max}=220$~MeV fit while it is given by the deviation of
the $E_{\rm max}=300$~MeV fit above 200 MeV. The second kink present
in the truncation error, on the other hand, is caused by the
transition of $Q$ from $M_\pi^{\rm eff}/\Lambda_{\rm b}$ to $p/\Lambda_{\rm b}$.

\begin{table}[t]
	\caption{Deuteron binding energy $B_d$, asymptotic S-state
          normalization $A_S$, asymptotic D/S-state ratio $\eta$,
          radius $r_d$,  quadrupole moment $Q$ and D-state
          probability $P_D$ as predicted by various high-quality
          potentials. The binding energy has been calculated with the
          nonrelativistic energy-momentum relation for the potentials
          of Ref.~\cite{Entem:2017gor} and with the relativistic
          relation for the SMS potential of
          Ref.~\cite{Reinert:2017usi} and the CD Bonn potential 
          Ref.~\cite{Machleidt:2000ge}. 
		\label{tab:deuteron-order-by-order}}
	\smallskip
	\begin{tabular*}{\textwidth}{@{\extracolsep{\fill}}lllllr}
	\hline 
	\hline
	\noalign{\smallskip}
	   &  \multicolumn{1}{c}{Granada} &
                                                                 \multicolumn{1}{c}{CD
                                            Bonn}  &
                                                     \multicolumn{1}{c}{EMN N$^4$LO$^+$ \cite{Entem:2017gor}}  & \multicolumn{1}{c}{SMS N$^4$LO$^+$ \cite{Reinert:2017usi}}  & Empirical
	\smallskip
	 \\
	  &\cite{Perez:2013mwa} & \cite{Machleidt:2000ge} &  \multicolumn{1}{c}{$\Lambda = 500$~MeV} & \multicolumn{1}{c}{$\Lambda = 450$~MeV}
	 \smallskip
	 \\
          \hline \hline
&&&&&\\[-7pt]          
	$B_d$ (MeV)         & 2.2246$^\star$ & 2.2246$^\star$ & 2.2246$^\star$ & 2.2246$^\star$ & 2.224575(9) \cite{VanDerLeun:1982bhg} \\ 
	$A_S$ (fm$^{-1/2}$) & 0.8829 & 0.8846 & 0.8852 & 0.8847 & 0.8846(8) \cite{Ericson:1982ei} \\ 
	$\eta$              & 0.0249 & 0.0256 & 0.0258 & 0.0255 & 0.0256(4) \cite{Rodning:1990zz} \\ 
	$r_d$ (fm)          & 1.965  & 1.966 &  1.973 & 1.966  & $1.97535(85)^\dagger$ \cite{Huber:1998zz}\\ 
	$Q$ (fm$^2$)        & 0.268  & 0.270 & 0.273  & 0.270  & 0.2859(3) \cite{Bishop:1979zz} \\ 
	$P_D$ ($\%$)        & 5.62   & 4.85  & 4.10   & 4.59   & --- \\[4pt]
	\hline \hline
	\end{tabular*}
	\begin{tabular*}{\textwidth}{@{\extracolsep{\fill}}l}
	$^\star$The deuteron binding energy has been
	  taken as input in the fit. \\[-2pt]
	$^\dagger$This value corresponds to the so-called deuteron
	structure radius, which is defined as a square root of \\[-1pt]
	the difference of the deuteron, proton and neutron mean square charge radii.
	\end{tabular*}
\end{table}
Tab.~\ref{tab:deuteron-order-by-order} shows the deuteron properties
as predicted by various high-quality potentials. 
Clearly, the error analysis can also be applied to the bound state
properties of Tab.~\ref{tab:deuteron-order-by-order}. However, the
obtained uncertainties are only meaningful for a complete
calculation of an unconstrained observable. This excludes the deuteron
binding energy $B_d$ (as it is a fitted quantity), the quadrupole
moment $Q$ and deuteron radius $r_d$ (as meson exchange currents and
relativistic corrections are not taken into account) as well as the
D-state probability $P_D$ (which is not observable). On the other
hand, we can perform the uncertainty quantification for the asymptotic
S-state normalization $A_S$ and the asymptotic D/S-state ratio $\eta$
for which we obtain at N$^4$LO$^+$ and for $\Lambda =450$~MeV the
values of $A_s = 0.8847_{(-3)}^{(+3)} (5) (0) (1)$ and $\eta =
0.02553_{(-9)}^{(+11)} (4) (3) (8)$, respectively. Here the first, second, third and
fourth error refer to the NN statistical, truncation, $\pi$N
statistical and $E_{\rm max}$ uncertainty, respectively. Notice that
the quoted truncation errors estimated using the Bayesian model of section
\ref{sec5} are fairly similar to the ones given in Ref.~\cite{Reinert:2017usi}, which
were obtained using the EKM method. On the other hand, the $\pi$N
statistical uncertainties are much smaller than the corresponding
errors quoted in Ref.~\cite{Reinert:2017usi}, where an attempt was made to also
include systematic effects by using the values of these LECs
determined in the physical region of $\pi$N scattering.

Finally, let us discuss the treatment of isospin-breaking effects in
the two-nucleon interaction. Like all modern high-precision
potentials, the SMS interactions include isospin-breaking in the OPEP
due to the different physical pion masses $M_{\pi^\pm}$ and
$M_{\pi^0}$ and charge dependence of the short-range potential in the
${}^1S_0$ partial wave. These are the dominant and well-understood
isospin-breaking effects necessary to arrive at e.g. a correct
description of the charge-dependence of the ${}^1S_0$ scattering
length. For the calculation of scattering observables in the
two-nucleon system, the isospin-breaking due to long-range
electromagnetic interactions is taken into account as discussed at the
beginning of this section. This treatment of strong and
electromagnetic isospin-breaking effects is identical to the Nijmegen
PWA \cite{Stoks:1993tb}. On the other hand, chiral EFT allows for a systematic inclusion
of isospin-breaking effects beyond the ones previously considered. In
fact, expressions for the leading isospin-breaking TPEP
\cite{Friar:1999zr,Friar:2003yv}, the subleading isospin-breaking TPEP
\cite{Epelbaum:2005fd} and irreducible $\pi\gamma$ exchange
\cite{vanKolck:1997fu}, which are (mostly) parameter-free in the
two-nucleon system, have been available for some time. The
long-standing question regarding the charge-dependence of the $\pi$NN
coupling constant also re-emerges in a systematic treatment of
isospin-breaking effects in the framework of chiral EFT. While the
Nijmegen group did not find evidence for charge-dependence, the issue
does not seem to be settled, see Ref.~\cite{Perez:2016aol} for a
recent determination. Last but not least, charge-dependence in the
short-range potential entering P-waves should also be taken into
account starting from N$^4$LO \cite{Epelbaum:2005fd}.

\subsection{Three-nucleon scattering}
\label{3Nresults}

As discussed in the previous subsection,  the N$^4$LO$^+$
SMS potentials of Ref.~\cite{Reinert:2017usi} lead to excellent and in fact
a nearly perfect description of np and pp scattering data
below pion production threshold. Moreover, an order-by-order comparison of the
results for various observables along with the Bayesian error analysis indicate a
generally good convergence of the chiral expansion in the NN
sector. On the other hand, a description of nucleon-deuteron
elastic and breakup scattering data at a comparable level of accuracy is not
available yet.
Extensive calculations performed in the last decades using
high-precision phenomenological NN potentials and 3NF models in the
framework of the Faddeev 
equations \cite{Gloeckle:1995jg}  and using other {\it ab initio} methods \cite{Kievsky:1998gt}
have revealed the following picture (see Ref.~\cite{KalantarNayestanaki:2011wz} and references therein):
\begin{itemize}
\item[--] Calculations based on high-precision NN potentials alone
  (including the N$^4$LO$^+$ ones of Ref.~\cite{Reinert:2017usi}) tend to
  underestimate the $^3$H and $^3$He binding energy by
  $\sim 0.5$~MeV and generally lead to similar predictions in Nd
  scattering observables. 
\item[--] At low energies, the resulting description of Nd data
  appears to be rather good apart from a few exceptions
  such as the underprediction of the nucleon
analyzing power $A_y$, known as the $A_y$  puzzle \cite{Witala:1991emr}, and the discrepancy
for the cross section for the symmetric space star deuteron breakup configuration
\cite{Witala:2009wd}. 3NF effects in this energy range are
found to be small
in agreement with qualitative arguments based on the chiral power
counting as explained below.  
\item[--] Starting from $E_{\rm N} \sim 50$~MeV, discrepancies between  theory
and experimental data set in and become large at $E_{\rm N} \sim 200$~MeV and
above. Except for the cross section, the inclusion of the
phenomenological 3NFs like the Tucson-Melbourne (TM99)  \cite{Coon:2001pv} and Urbana-IX
\cite{Pudliner:1995wk} models does not globally reduce the discrepancies
between theory and data
\cite{KalantarNayestanaki:2011wz}. Relativistic effects have also been
studied, see Ref.~\cite{Witala:2011yq} and references therein, and found to
be small at energies below the pion
production threshold. 
\end{itemize}
Assuming that the discrepancies between theory and experimental data
in the 3N system are to be resolved by 3NFs, these findings demonstrate
that the currently available phenomenological models do not provide an
appropriate description of the 3NF.  This should not
come as a surprise given the enormously rich and complex spin-isospin-momentum
structure of a most general 3NF \cite{Krebs:2013kha,Phillips:2013rsa,Epelbaum:2014sea,Topolnicki:2017rnt}. 
Here, chiral EFT offers a decisive advantage over more
phenomenological approaches by predicting the long-range
part of the 3NF in a model-independent way, establishing a clear importance
hierarchy of short-range terms and providing a solid theoretical
framework for maintaining consistency between two- and three-nucleon
forces and ensuring scheme independence of the calculated observables.  

As already mentioned in section~\ref{sec3}, three-body contributions to
the nuclear Hamiltonian first appear at N$^2$LO in the chiral
expansion and are, therefore, suppressed by $Q^3$ relative to the
dominant pairwise NN interaction. It is instructive to estimate the
expected magnitude of 3NF effects for various observables solely on
the basis of the chiral 
power counting (i.e.~using NDA). For $^3$H and $^4$He, one can use the
typical expectation values of the NN potential energy of
$\langle V_{\rm NN} \rangle_{\rm ^3 H} \sim 50$~MeV and $\langle V_{\rm
  NN} \rangle_{\rm ^4 He} \sim 100$~MeV \cite{Binder:2015mbz}, along with the estimation
of the expansion parameter $Q\sim M_\pi^{\rm eff} / \Lambda_{\rm b}$ with
$M_\pi^{\rm eff} = 200$~MeV and  $\Lambda_{\rm b} = 650$~MeV, in order to estimate the
expected 3NF contributions to the binding energy to be $\langle V_{\rm
  3N} \rangle_{\rm ^3 H} \sim Q^3 \langle V_{\rm NN} \rangle_{\rm ^3 H} \sim
1.5$~MeV and $\langle V_{\rm
  3N} \rangle_{\rm ^4 He} \sim Q^3 \langle V_{\rm NN} \rangle_{\rm ^4 He} \sim
3$~MeV. These qualitative estimations agree well with the actual
underprediction of the
$^3$H and $^4$He by the NN interactions alone which, using the AV18
\cite{Wiringa:1994wb}, CD Bonn \cite{Machleidt:2000ge}, N$^2$LO
\cite{Epelbaum:2004fk} and Idaho N$^3$LO \cite{Entem:2003ft} potentials
as representative examples, 
amounts to $0.5\ldots 0.9$~MeV and $2.1\ldots 4.1$~MeV, respectively. 
The shallow nature of few-nucleon bound states indicates that there
are large cancellations between the kinetic and potential energies. Because of
this fine tuning, 3NF contributions to the binding energies are
enhanced beyond the naive estimation of $Q^3 \sim 3$\% and actually reach $10 \ldots 15$\%.    
On the other hand, there is generally no reason to expect a similar
enhancement for Nd scattering observables at low energy except for
some fine-tuned polarization observables such as $A_y$. It is
well known that tiny changes of the NN interaction in the
triplet $P$-waves amount to large relative changes in the Nd $A_y$
\cite{Witala:1991emr}. On the other hand, starting from $E_{\rm N} \sim 60$~MeV, the
expansion parameter $Q$ in Eq.~(\ref{ExpPar}) is dominated by the 
momentum scale $p = \sqrt{2/3 m E_{\rm N}}$ \cite{Epelbaum:2019zqc}. At  
e.g.~the energies of $E_{\rm N} \sim 100$~MeV and $E_{\rm N} \sim 200$~MeV,
the expansion parameter becomes $Q \sim 0.40$ and  $Q \sim 0.55$, and
the relative contributions of the 3NF to a generic scattering observable are expected
to increase to $\sim 6$\% and   $\sim 16$\%, respectively.  Clearly,
these simplistic back-of-envelope estimations only yield qualitative
insights into the role of the 3NF. Nevertheless, they
agree remarkably well with the observed trend of discrepancies between
theoretical predictions based solely on the NN interactions and
experimental data, which tend to increase with energy. For further
examples and a more quantitative
analysis along this line of Nd scattering, selected properties of
light and medium-mass nuclei and the equation of state of nuclear
matter see Refs.~\cite{Binder:2015mbz,Binder:2018pgl,Hu:2016nkw,Hu:2019zwa}. We further emphasize that it
is not entirely clear how to estimate the relevant momentum scale,
that determines the expansion parameter in heavy
nuclei, and how to quantify truncation errors for excited states, see
Ref.~\cite{Binder:2018pgl} for an extended discussion. 

As discussed in section~\ref{sec3} and visualized in Fig.~\ref{chiralexpansionNN}, the leading contributions to the 3NF
at N$^2$LO emerge from the two-pion exchange,
one-pion-exchange-contact and purely contact tree-level diagrams,
leading to the well-known expressions \cite{vanKolck:1994yi,Epelbaum:2002vt}
\beqa
\label{leading}
V_{\rm 3N} &=& \frac{g_A^2}{8 F_\pi^4}\; 
\frac{\vec \sigma_1 \cdot \vec q_1  \; \vec \sigma_3 \cdot \vec q_3 }{(
 \vec  q_1^{\, 2} + M_\pi^2) \, (\vec q_3^{\, 2} + M_\pi^2)} \;\Big[ \fet \tau_1
\cdot \fet \tau_3  \big( - 4 c_1 M_\pi^2 
+ 2 c_3 \, \vec q_1 \cdot \vec
    q_3 \big)  
+  c_4 \fet \tau_1
  \times \fet \tau_3  \cdot \fet \tau_2  \; \vec q_1 \times \vec q_3 
\cdot \vec \sigma_2  \Big]  \nn
&-& \frac{g_A \, D}{8 F_\pi^2}\;  
\frac{\vec \sigma_3 \cdot \vec q_3 }{\vec q_3^{\, 2} + M_\pi^2} \; 
\fet \tau_1 \cdot \fet \tau_3 \; \vec \sigma_1 \cdot \vec q_3 \; + \; 
\frac{1}{2} E\, \fet \tau_1 \cdot \fet \tau_2
\; + \;  \mbox{5 permutations}\,, 
\eeqa
where  $\vec q_{i} = \vec p_i \,
' - \vec p_i$  with $\vec p_i \, '$
and $\vec p_i$ being the final and initial momenta of the nucleon $i$.  
The LECs $D$ and $E$ are usually expressed 
in terms of the corresponding dimensionless
coefficients $c_D$ and $c_E$ via $D= c_D/(F_\pi^2 \Lambda_\chi )$
and $E= c_E/(F_\pi^4 \Lambda_\chi)$ \cite{Epelbaum:2002vt}.
In Refs.~\cite{Epelbaum:2018ogq} and \cite{Epelbaum:2019zqc},
semilocal coordinate- and momentum-space regularized 3NF expressions
in combination with the corresponding chiral NN potentials from Refs.~\cite{Epelbaum:2014efa,Epelbaum:2014sza}
and \cite{Reinert:2017usi}, respectively, were employed by the LENPIC Collaboration
to analyze Nd scattering observables at N$^2$LO.  The numerical
implementation of the 3NF in the Faddeev equations is carried out in
the partial wave basis. Partial wave decomposition (PWD) of a general
3NF can be carried out numerically using the machinery developed in
Ref.~\cite{Golak:2009ri} by performing five-dimensional
angular integrations. Given the required number of partial waves
and grid points for the four Jacobi momenta to reach converged results
for Nd scattering observables, such a numerical PWD requires
substantial computational resources. In Ref.~\cite{Hebeler:2015wxa}, a more efficient
approach was introduced, that exploits the local nature of the bulk of
the 3NF.    

To make predictions for few-nucleon observables, one first
needs to determine the LECs $c_D$ and $c_E$ entering the 3NF. A broad
range of few- and many-body observables including the binding energies
and radii of $^3$H, $^4$He and heavier nuclei, nucleon-deuteron
doublet scattering length $^2a$, n-$\alpha$ scattering, triton $\beta$-decay
and the saturation properties of nuclear matter
have been proposed and employed in the past to determine these two LECs
\cite{Epelbaum:2002vt,Navratil:2007we,Gazit:2008ma,Piarulli:2017dwd,Lynn:2017fxg,Ekstrom:2015rta}.
A reliable determination of $c_D$, $c_E$ is complicated by the existence
of strong correlations between some of the low-energy observables,
see e.g.~\cite{Platter:2004zs}, which originate from the large S-wave scattering lengths in the
NN system. Furthermore, going beyond the 3N system may
require, depending on the observable and the chiral order, the
inclusion of  4NF and exchange current contributions. In Ref.~\cite{Epelbaum:2018ogq} we,
therefore, restricted ourselves to 3N observables in the determination
of $c_D$, $c_E$.
Specifically, we employed the $^3$H binding energy of $B_{\rm
  ^3H} = 8.482$~MeV to fix the LECs $c_E$ for a given value of $c_D$. 
The remaining LEC $c_D$ was determined
by considering a
number of observables including
$^2a = 0.645 \pm 0.008$~fm \cite{Schoen:2003my}, nd total cross section data from \cite{Abfalterer:2001gw} and
precisely measured pd
differential cross section in the minimum region at $E_{\rm N} =
70$~MeV \cite{Sekiguchi:2002sf}, $108$~MeV \cite{Ermisch:2003zq} and $135$~MeV \cite{Sekiguchi:2002sf}. 
\begin{figure}[tb]
\begin{center}
\includegraphics[width=\textwidth]{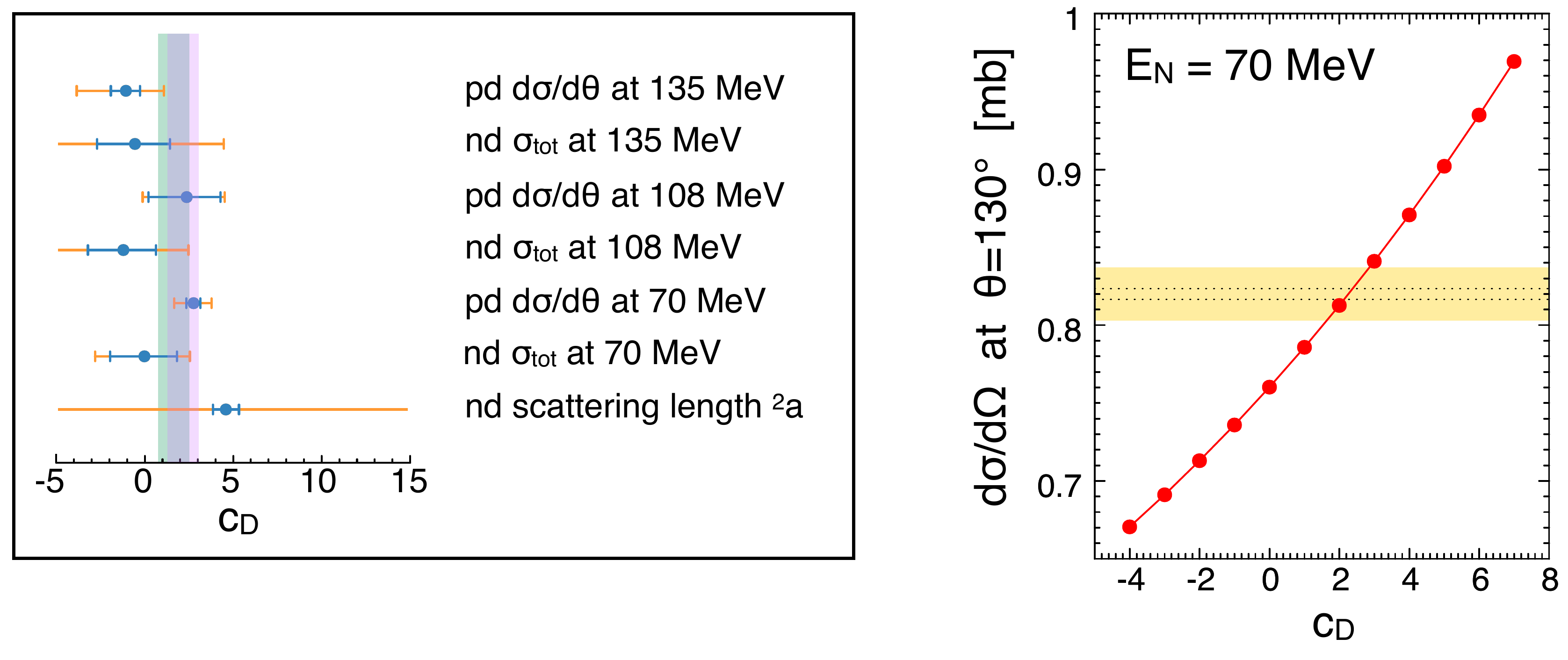}
\end{center}
\vskip -0.6 true cm 
\caption{Left: Determination of the LEC $c_D$ at N$^2$LO from selected
  Nd
  scattering observables.  The smaller (blue) error bars
  correspond to the experimental uncertainty while the larger (orange)
  error bars also take into account the truncation error at
  N$^2$LO estimated using the EKM approach of
  Ref.~\cite{Epelbaum:2014efa}. The green (violet) bands show standard
  error intervals of $c_D$ resulting from a combined least
  squares single-parameter fit to all observables (to observables up to
  $E_{\rm N}  =108$~MeV) using the orange error bars. 
  Right: Nd cross section in the minimum region
  ($\theta = 130^\circ$) at $E_N = 70$~MeV as function of the LEC $c_D$. For
  each $c_D$ value, the LEC $c_E$ is adjusted to the $^3$H binding
  energy. Dotted lines show the statistical uncertainty of the
  experimental data from Ref.~\cite{Sekiguchi:2002sf}, while the yellow band also
  takes into account the quoted systematic uncertainty of $2\%$.
All results are obtained using the N$^2$LO SCS NN potential from Ref.~\cite{Epelbaum:2014efa}
  in combination with the N$^2$LO SCS 3NF for the
  coordinate-space cutoff $R=0.9$~fm. \label{fig:cDcE} }
\end{figure}
In the left panel of Fig.~\ref{fig:cDcE}, we show the extracted values
of $c_D$ for the SCS  interactions
with the cutoff $R = 0.9$~fm. It is reassuring to see
that  the
considered 3N observables lead to consistent values of $c_D$.  In
addition, these results show that the
strongest constraint on $c_D$, given the experimental \emph{and
the estimated truncation uncertainty}, is imposed by the pd differential
cross section data at   $E_{\rm N} =
70$~MeV from Ref.~\cite{Sekiguchi:2002sf} as visualized in the right
panel of Fig.~\ref{fig:cDcE}. We also found no correlations between
this observable and the $^3$H binding energy. In particular, the
resulting value of the LEC $c_D$ is largely determined by the
differential cross section and
almost insensitive to a variation of the triton
binding energy.  

\begin{figure}[tb]
\begin{center}
\includegraphics[width=\textwidth]{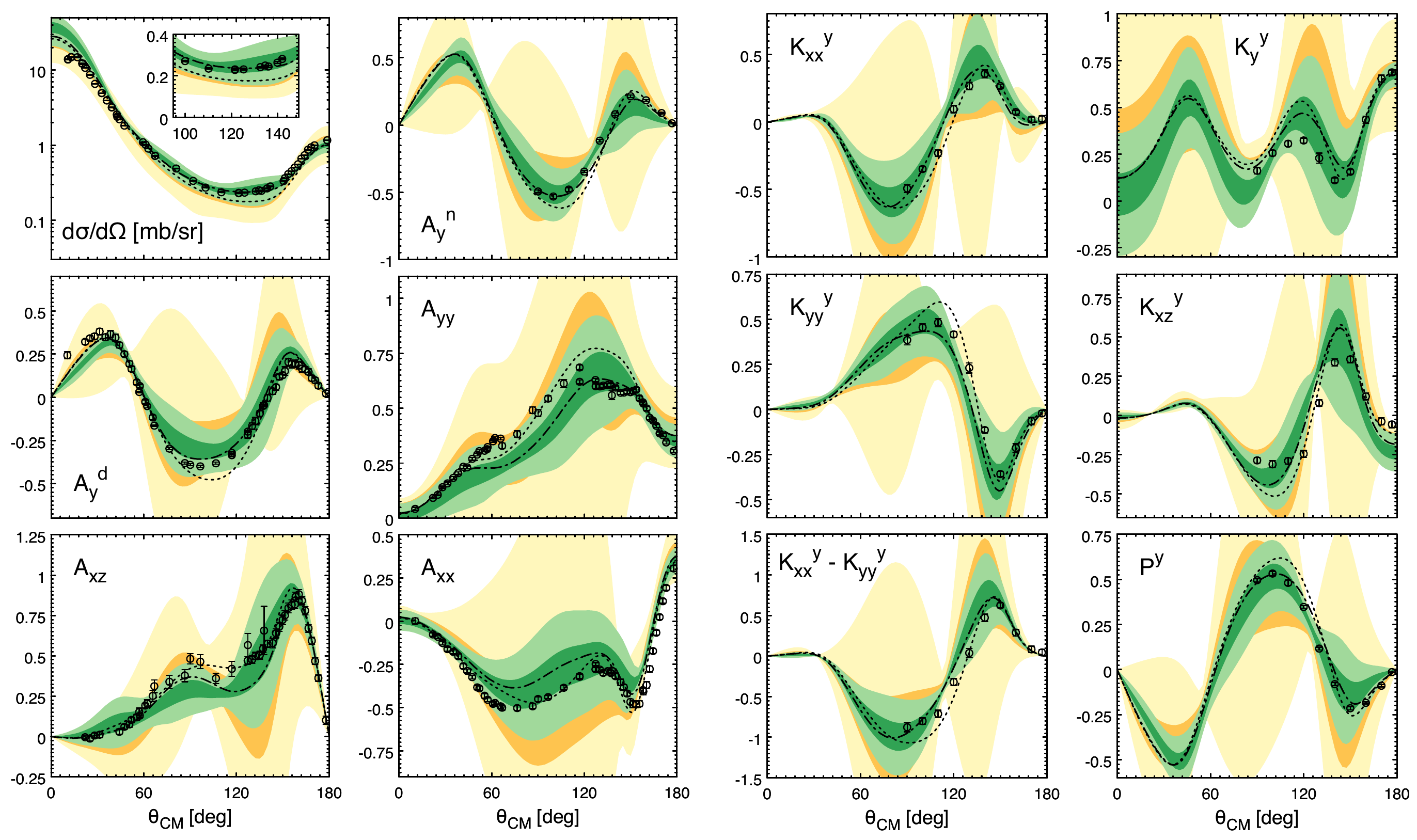}
\end{center}
\caption{Predictions for the differential cross section, nucleon and
  deuteron analyzing powers $A_y^n$ and $A_y^d$, deuteron
  tensor analyzing powers $A_{yy}$, $A_{xz}$, $A_{xx}$,  polarization transfer coefficients
$K_{xx}^y$, $K_{y}^y$, $K_{yy}^y$, $K_{xz}^y$, $K_{xx}^y-K_{yy}^y$
  and the induced polarization $P^y$
  in elastic
  Nd scattering  at laboratory energy of
  $E_{\rm N} = 135$~MeV at NLO
  (yellow bands) and N$^2$LO (green bands).
  The light- (dark-) shaded bands indicate $95\%$ ($68\%$)
DoB intervals using the Bayesian model $\bar C_{0.5-10}^{650}$
introduced in section~\ref{sec5}. 
  Open circles are proton-deuteron data from
  Ref.~\cite{Sekiguchi:2002sf}.  The dotted (dashed-dotted) lines show the results based on the
CD Bonn NN  potential \cite{Machleidt:2000ge} (CD Bonn NN potential in combination with
the  Tucson-Melbourne 3NF \cite{Coon:2001pv}). All results are obtained using the N$^2$LO SMS NN potential from Ref.~\cite{Reinert:2017usi}
  in combination with the N$^2$LO SMS 3NF for the
  momentum-space cutoff $\Lambda = 500$~MeV. 
}
\label{fig:135MeV}
\end{figure}

In Ref.~\cite{Epelbaum:2019zqc}, we have analyzed Nd scattering observables 
using the most recent SMS  NN
potentials from Ref.~\cite{Reinert:2017usi} in combination with the N$^2$LO 3NF
regularized in the same way. Motivated by the experience with the SCS
interactions \cite{Epelbaum:2018ogq}, the
LECs $c_D$ and $c_E$ were determined from the $^3$H binding energy and
the pd cross section minimum at $E_{\rm N} = 70$~MeV. In Fig.~\ref{fig:135MeV}
we show, as a representative example, our N$^2$LO predictions for selected Nd scattering observables
at $E_{\rm N} = 135$~MeV, 
along with the experimental data and calculations based on the CD
Bonn NN potential with and without the TM99 3NF model. 
As an important internal consistency check of the calculations, we have
verified that the predictions obtained using different cutoff values
are consistent with each other (within errors), see Fig.~5 of
Ref.~\cite{Epelbaum:2019zqc}. 

It is reassuring to see that the experimental data are generally well described
by the theory. On the other hand, while accurate, our predictions at 
N$^2$LO have obviously rather low precision at this energy. In fact,
the  N$^2$LO truncation errors are
comparable with or even larger than the observed deviations between experimental data and
calculations based on phenomenological high-precision NN and 3NF
models, see the dotted and dashed-dotted lines in Fig.~\ref{fig:135MeV}.    
Based on the experience in the NN sector as discussed in section
\ref{NNresults}, it is conceivable that a high-precision description of Nd scattering
data will require the chiral expansion of the 3NF to be pushed to (at least)
N$^4$LO. At the energy of $E_{\rm N} = 135$~MeV, the uncertainty bands at
N$^4$LO are expected to become $4$-$5$ times more narrow as compared with
the N$^2$LO ones shown in Fig.~\ref{fig:135MeV}.  

So where do we stand in terms of efforts to include 3NF corrections
beyond N$^2$LO? As explained in section~\ref{subsec:Reg}, the main obstacle for
the inclusion of higher order contributions to the 3NF is the lack of
their \emph{consistently regularized} expressions. Starting from N$^3$LO, it
is not sufficient anymore to naively regularize the available
expressions for the 3NF from
Refs.~\cite{Bernard:2007sp,Bernard:2011zr,Krebs:2012yv,Krebs:2013kha}
derived using DR, since such an approach violates constraints
imposed by the chiral symmetry. Rather, the N$^3$LO and N$^4$LO corrections to
the 3NF need to be re-derived using the consistent finite-cutoff
regularization approach. Work along these lines is in progress.
Another challenge, that will have to be addressed at N$^4$LO, is the
determination of the LECs appearing in the 3NF at this order. While
the N$^3$LO contributions do not involve free parameters, the
short-range part of the 3NF at N$^4$LO depends on $10$ unknown
LECs \cite{Girlanda:2011fh}, from which $9$ contribute to the isospin-$1/2$ channel and thus
can, in principle, be determined in Nd scattering. Furthermore, the
yet-to-be-derived  one-pion-exchange-contact contributions to the 3NF
at N$^4$LO will also involve unknown LECs. Given the still rather
significant computational cost of solving the Faddeev equations in the
3N continuum, the complicated treatment of the Coulomb interaction
\cite{Deltuva:2005wx} and the lack of partial wave analyses in the 3N
sector,
the determination of
these LECs from 3N scattering data will certainly be a challenging
task. 

While a complete analysis of Nd scattering is currently not available
beyond N$^2$LO, it is instructive to explore the role of
subleading short-range 3NF interactions. In Ref.~\cite{Girlanda:2018xrw}, it was shown
within a hybrid phenomenological approach
that the 3N contact operators at N$^4$LO can be tuned to reproduce the $^3$H
binding energy, nd scattering lengths, cross section and polarization
observables of pd scattering at $2$~MeV center-of-mass energy. The
resulting models were shown to lead to a satisfactory description of
pd polarization observables below the deuteron breakup. On the other
hand, 3NF effects are expected to be much more visible at intermediate
and higher energies. In Ref.~\cite{Epelbaum:2019zqc}, we explored the impact of the
short-range 3NF operators of the central and spin-orbit types
proportional to the LECs $E_1$ and $E_7$, respectively,  
\beq
 V_{3N} = E_1 \,  \vec q_1^{\, 2} \; + \; i E_7 \, \vec q_1 \times (\vec K_1 - \vec K_2)
 \cdot (\vec\sigma_1 + \vec{\sigma_2})
\;  + \;  \mbox{5 permutations}\,, 
 \eeq
 where $\vec K_i = (\vec p_i \, ' + \vec p_i )/2$. Parametrizing the
 dimension-full LECs $E_1$, $E_7$ in terms of the corresponding
 dimensionless parameters via $E_i = c_{E_i} /(F_\pi^4 \Lambda_\chi^3
 )$ with $\Lambda_\chi = 700$~MeV, we studied the impact of these
 N$^4$LO terms on selected Nd scattering observables for the fixed
 values of the LECs of $c_{E_i} = \pm 2$. Based on the experience in
 the NN sector and with the N$^2$LO 3NF, we expect the actual values
 of these LECs to lie well within this range.
The expectation values of various contributions to the 3NF  in the triton
 state indicate that the employed values $c_{E_7} = \pm 2$ may already
  overestimate the expected natural range of this LEC.

\begin{figure}[tb]
\begin{center}
\includegraphics[width=\textwidth]{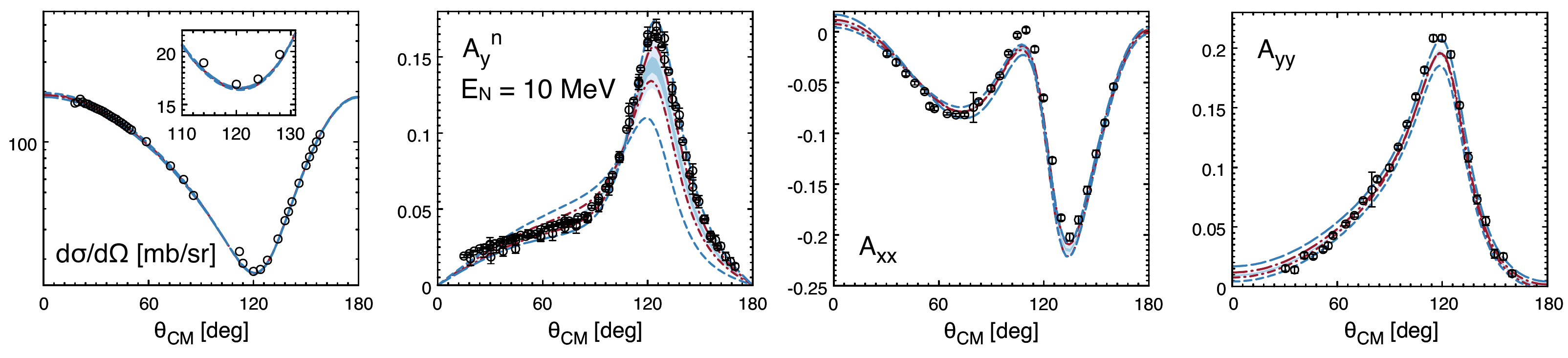}
\end{center}
\caption{Results for the differential cross section, nucleon analyzing
  powers $A_y^n$ as well as deuteron
  tensor analyzing powers $A_{xx}$ and $A_{xx}$ in elastic
  nucleon-deuteron scattering  at laboratory energy of
  $E_{\rm lab}^N = 10$~MeV based on the SMS NN
  potentials of Ref.~\cite{Reinert:2017usi} at N$^4$LO$^+$ in
  combination  with the SMS 3NF at
  N$^2$LO using $\Lambda = 450$~MeV. Blue light- (dark-) shaded bands 
 show the expected truncation
  uncertainty for a \emph{complete} N$^3$LO calculation and are
  obtained by multiplying the N$^2$LO truncation error corresponding to
  $95\%$ ($68\%$) DoB intervals for the
  model $\bar C_{0.5-10}^{650}$ with the corresponding value of the expansion parameter
  $Q$. Short-dashed-dotted and long-dashed-dotted red lines show the impact
  of the N$^4$LO central short-range 3NF $\propto c_{E_1}$ with $c_{E_1} = -2$ and
  $c_{E_1} = 2$, respectively.  Similarly, short-dashed and
  long-dashed blue lines show the impact
  of the N$^4$LO spin-orbit short-range 3NF $\propto c_{E_7}$ with $c_{E_7} = -2$ and
  $c_{E_7} = 2$, respectively.  Open circles are
neutron-deuteron data from Ref.~\cite{data10_4} and proton-deuteron data from
Ref.~\cite{data10_1,data10_2,data10_3}, corrected for the Coulomb effects, see Ref.~\cite{Epelbaum:2002vt} for
details.  
}
\label{fig:E1E7_10MeV}
\end{figure}

 In order to compute the contributions of the $c_{E_i}$-terms to 3N observables
 in a meaningful way, one needs to perform (implicit) renormalization
 as explained in section~\ref{sec2}. This was achieved in Ref.~\cite{Epelbaum:2019zqc} by
simultaneously adjusting the values of the N$^2$LO LECs $c_D$, $c_E$ to the triton
binding energy and the cross section minimum at $E_{\rm N} = 70$~MeV
for all considered values of the  LECs $c_{E_i}$. The calculations have
been performed using the N$^4$LO$^+$  SMS NN potential from
Ref.~\cite{Reinert:2017usi} in combination with the SMS N$^2$LO 3NF. 
In Fig.~\ref{fig:E1E7_10MeV}, we show the resulting predictions at the
lowest considered energy of $E_{\rm N} = 10$~MeV. 
The blue bands show the estimated truncation error at N$^3$LO, 
obtained by rescaling the N$^2$LO Bayesian truncation uncertainty with
the expansion parameter $Q$ \footnote{We cannot estimate the
  N$^3$LO truncation error using the Bayesian approach described in
  section~\ref{sec5} since no complete N$^3$LO results are available for
  Nd scattering.}, and visualize the expected impact of N$^4$LO terms.   
In agreement with the expectations, 3NF effects generally appear to be
rather small at such low energies. This figure also provides a nice
illustration of the fine tuned nature of the nucleon vector analyzing
power $A_y$, which shows a strong sensitivity to small changes in the
Hamiltonian. What has
been traditionally referred to as the  $A_y$-puzzle thus appears to be
just a consequence of the fine-tuned nature of this observable, and the
``puzzle'' may be expected to be resolved by 3NF contributions beyond
N$^2$LO.   
While the $A_y$ is well known to be particularly sensitive to
spin-orbit types of 3NFs \cite{Kievsky:1999nw} such as the one proportional to
$c_{E_7}$, our results also show an unexpectedly strong sensitivity to
the subleading central interaction of the $c_{E_1}$-type. 

\begin{figure}[tb]
\begin{center}
\includegraphics[width=\textwidth]{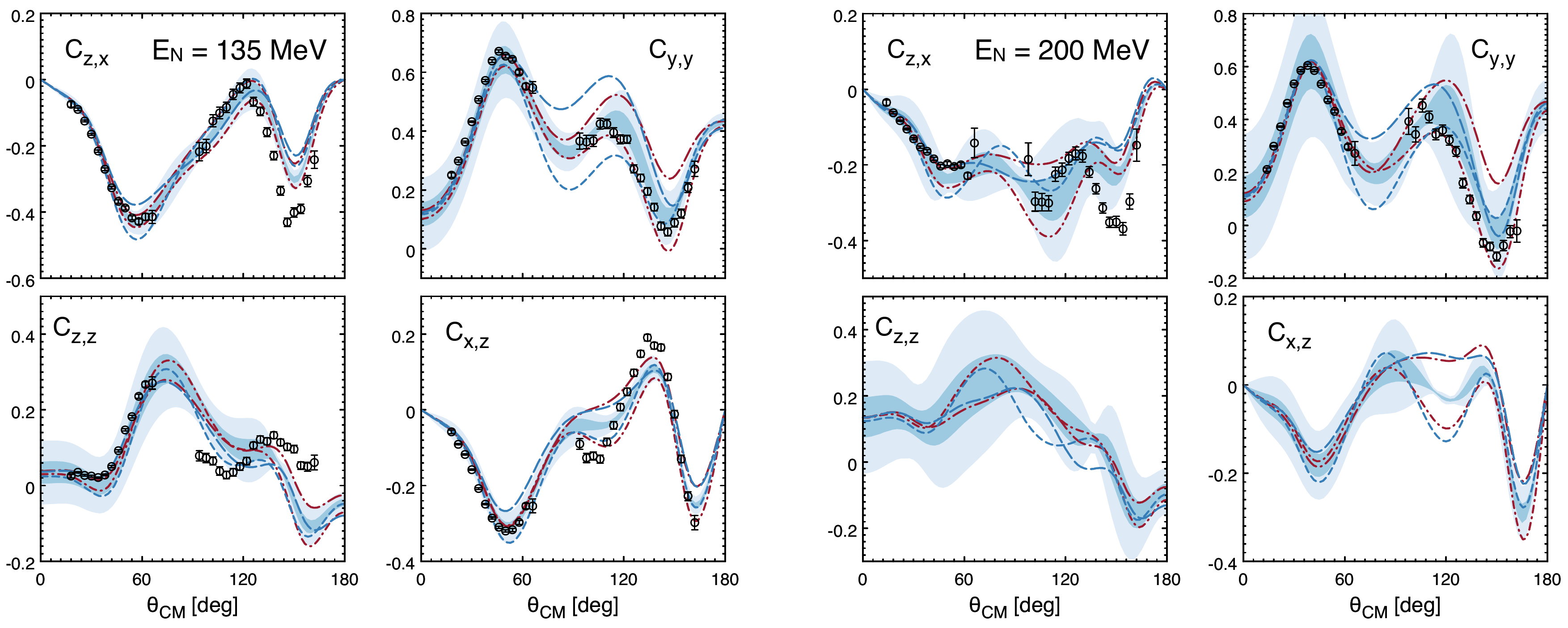}
\end{center}
\caption{
  Same as Fig.~\ref{fig:E1E7_10MeV} but for the
  deuteron-nucleon spin-correlation 
  parameters $C_{z,x}$, $C_{y,y}$, $C_{z,z}$ and $C_{x,z}$ for Nd
  elastic scattering at $E_{\rm
    N} = 135$~MeV (left panel) and $E_{\rm
    N} = 200$~MeV (right panel).  Open circles are
 proton-deuteron data from
Ref.~\cite{vonPrzewoski:2003ig}. 
}
\label{fig:E1E7_135_200MeV}
\end{figure}

At higher energies, the effects of the considered N$^4$LO 3NF terms
become more significant as visualized in
Fig.~\ref{fig:E1E7_135_200MeV} for the case of selected
spin-correlation parameters. 
More results for the cross section, vector and tensor analyzing powers
and polarization transfer coefficients at $E_{\rm N} = 135$~MeV can be
found in Ref.~\cite{Epelbaum:2019zqc}. It is comforting to see that
the impact of the $c_{E_i}$-terms on Nd scattering observables is,
in general, consistent with the estimated N$^3$LO truncation
errors. One should, however, keep in mind that the employed Bayesian
approach may, under certain circumstances, become unreliable. This is,
in particular, the case for observables that depend on a continuously
varying parameter in the kinematical regions where the LO results and
higher-order corrections change sign,
see  Ref.~\cite{Epelbaum:2019zqc} for a detailed discussion.  One such
failure of the Bayesian model is shown  in
Fig.~\ref{fig:E1E7_135_200MeV} for the spin-correlation
coefficient $C_{x,z}$ at $E_{\rm N} = 200$~MeV around $\theta =
120^{\circ}$.  In such problematic cases, the approach proposed in
Ref.~\cite{Melendez:2019izc} and based on Gaussian processes is expected to provide more
reliable estimations of the truncation uncertainty. 

\subsection{Light nuclei}

While no results for light nuclei using SMS chiral interactions 
are available yet, we briefly review here some recent highlights 
obtained by the LENPIC Collaboration using the SCS NN potentials of
Refs.~\cite{Epelbaum:2014efa,Epelbaum:2014sza} with and without the
corresponding 3NFs at N$^2$LO.   In
Refs.~\cite{Binder:2015mbz,Binder:2018pgl}, we have calculated the
ground state energies and selected properties of  
light and medium-mass nuclei up to $^{48}$Ca using the
SCS NN interactions at various chiral orders.
Specifically, $A=3,4$ nuclei were analyzed in the framework of the
Faddeev-Yakubovsky equations while light p-shell nuclei were
calculated using the No-Core Configuration Interaction (NCCI) method
\cite{Barrett:2013nh,Maris:2008ax,Maris:2013poa} and employing Similarity Renormalization Group (SRG)
transformed interactions \cite{Glazek:1993rc,Wegner:1994,Bogner:2007rx,Bogner:2009bt} to improve the convergence.  
The results for $^{16,24}$O and $^{40,48}$Ca were obtained within the
coupled cluster and in-medium SRG group frameworks, see Refs.~\cite{Hagen:2007ew,Roth:2011vt,Binder:2013oea, Gebrerufael:2016xih, Stroberg:2016ung} and
references therein. A qualitatively similar convergence pattern was
observed in all considered cases, namely a significant overbinding at
LO, results close to the experimental values at NLO and N$^2$LO and
underbinding at N$^3$LO and N$^4$LO. Notice that the strongly
repulsive nature of the N$^3$LO contributions to the SCS NN interactions of
\cite{Epelbaum:2014efa,Epelbaum:2014sza} was shown
 to be caused by the employed unnaturally large
values of the redundant short-range operators \cite{Reinert:2017usi}. The SMS interactions of
Ref.~\cite{Reinert:2017usi} utilize a soft choice for these contact
terms, which leads to
more perturbative interactions at and beyond N$^3$LO.
No large gap between the N$^2$LO and N$^3$LO results for
the ground state energies is, therefore, expected for the new SMS NN interactions. 
The calculated charge radii of the considered medium-mass nuclei were
found to show a systematic improvement with the chiral order, but remain
underestimated using the NN interaction at the highest available order
N$^4$LO$^+$. 

\begin{figure}[tb]
\begin{center}
\includegraphics[width=\textwidth]{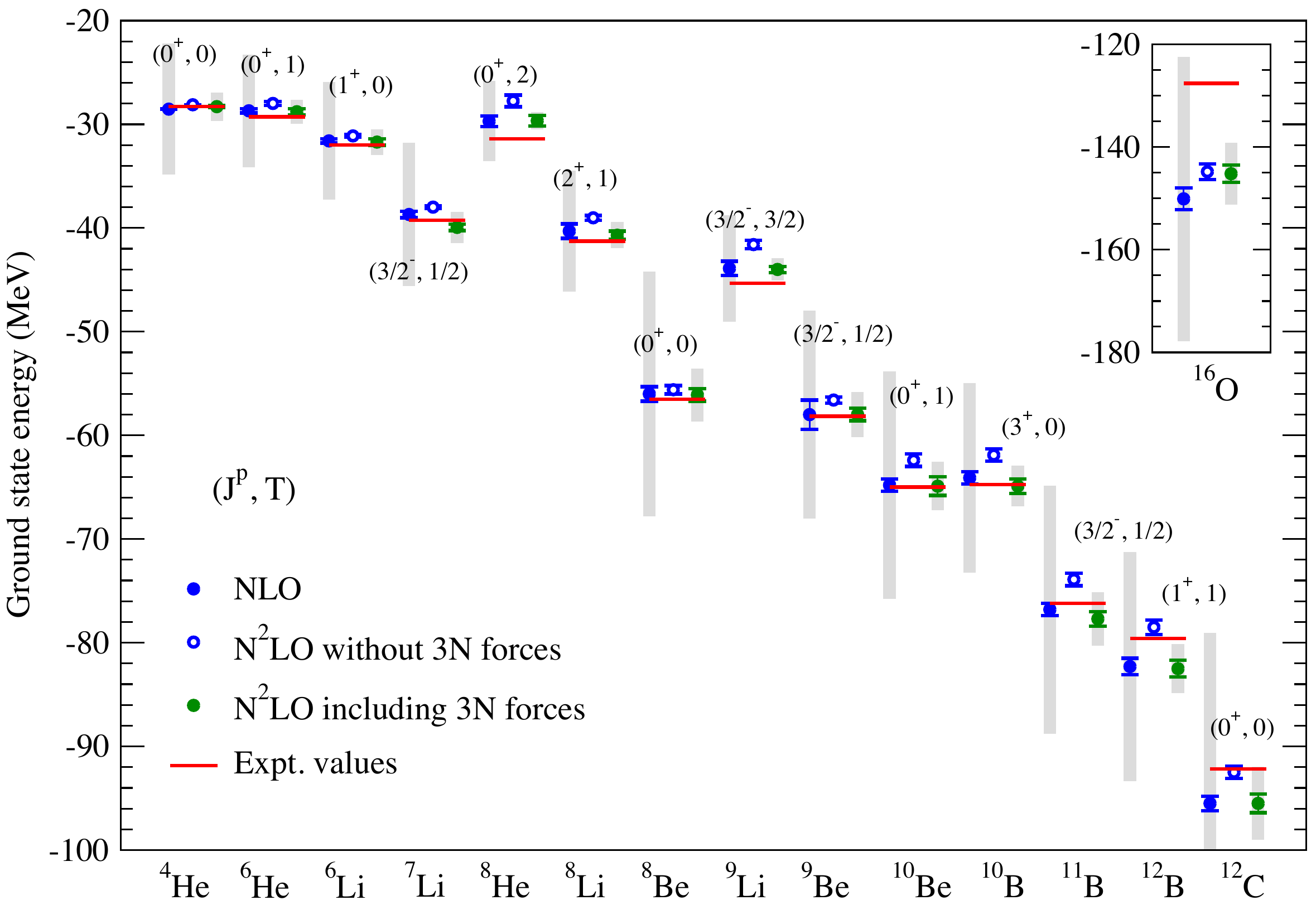}
\end{center}
\caption{Calculated ground state energies in MeV using chiral
SCS NN interactions from Ref.~\cite{Epelbaum:2018ogq}
in combination with the SCS 3NF at $R=1.0$~fm (open and solid dots) in
  comparison with experimental values (red levels).  For each nucleus
  the NLO and N$^2$LO results  are the left and right symbols
  and bars, respectively. The open blue
  symbols correspond to incomplete calculations at N$^2$LO using
  NN-only interactions.  Blue and green error bars indicate the NCCI
  extrapolation uncertainty and, where applicable, an estimate of the
  SRG dependence.  The shaded bars indicate the  truncation
  error at each chiral order corresponding to $68$\% DoB intervals
  using the Bayesian model $\bar C_{0.5-10}^{650}$ with the expansion
  parameter $Q = M_\pi^{\rm eff}/\Lambda_{\rm b}$.  
}
\label{fig:Spectra}
\end{figure}

In Ref.~\cite{Epelbaum:2018ogq}, a \emph{complete} N$^2$LO analysis of p-shell nuclei
was presented by the LENPIC Collaboration
using the SCS NN and 3N interactions. In Fig.~\ref{fig:Spectra},
we show the NLO and N$^2$LO results from that paper for nuclei up
to $A =16$.  
We emphasize that since the Hamiltonian has been completely determined
in the NN and 3N system as described in sections \ref{NNresults} and \ref{3Nresults},
the ground-state energies shown in that figure are parameter-free
predictions. In Fig.~\ref{fig:Spectra}, we have updated the
corresponding figure from Ref.~\cite{Epelbaum:2018ogq} by replacing the
truncation errors, that have been estimated in that paper using the EKM approach of 
Refs.~\cite{Epelbaum:2014efa,Binder:2018pgl}, with the Bayesian
uncertainties calculated as described in section~\ref{sec5}. The
$68$\% DoB Bayesian truncation
errors are similar to those quoted in Ref.~\cite{Epelbaum:2018ogq} at
N$^2$LO but appear to be  
significantly larger at NLO. We also calculated in that paper the
excitation energies for selected states of $A = 6-12$ nuclei. 
For almost
all considered cases, adding the 3NF to the NN interaction was found
to lead to a significant improvement in the description of experimental data.
The predicted ground state energies of p-shell nuclei show a good
agreement with the data except for $^{16}$O, which appears to be
overbound. Notice that the deviation between the predicted and
experimental values of the $^{16}$O binding energy  agrees with the
$95$\% DoB Bayesian error at N$^2$LO.
It will be very interesting to repeat the calculations for
the newest SMS interactions and to extend them to higher orders.

\section{Summary and outlook}
\label{sec7}

In this review article we have presented a snapshot of the current
state-of-the-art in low-energy nuclear theory with a focus on
the latest generation of semilocal nuclear potentials from  chiral EFT.
We now summarize some of the key conclusions of our
paper. 
\begin{itemize}
\item
  We have presented a concise and self-contained introduction to
  the conceptual foundations of
  chiral effective field theory
  in the few-nucleon sector and described in some detail all steps
  needed to compute low-energy observables from the effective chiral
  Lagrangians (including error analysis). Special emphasis was given to clarify the notion of
  \emph{consistency} of nuclear forces and current operators in terms of a 
  perturbative matching to the unambiguously defined on-shell
  scattering amplitude. In particular, few-nucleon potentials from
  Refs.~\cite{Epelbaum:2007us,Bernard:2007sp,Bernard:2011zr,Krebs:2012yv,Krebs:2013kha}
  and 
  electroweak current operators from
  Refs.~\cite{Kolling:2009iq,Kolling:2011mt,Krebs:2016rqz,Krebs:2019aka}
  at N$^3$LO and beyond, derived using 
  DR, are off-shell consistent with each other
  provided DR is also used to compute loop integrals arising from
  iterations of the dynamical equations. 
\item
  We have reviewed the semilocal momentum-space regularized potentials of
  Ref.~\cite{Reinert:2017usi}, which are currently the most precise chiral EFT NN forces
  on the market. These are the only
  NN interactions derived in chiral EFT, which -- from the statistical
  point of view -- qualify to be regarded as PWA of NN data below pion
  production threshold, see section~\ref{NNresults}
  for details. At the highest considered order N$^4$LO$^+$, these
  interactions describe the np and pp data from the
  self-consistent Granada-2013 database with a precision that is at least comparable to
  the one reached by modern phenomenological potentials
  with a much larger number of adjustable parameters. The significantly better description of the scattering
  data by the SMS N$^4$LO$^+$ interactions of Ref.~\cite{Reinert:2017usi} as compared to the
  nonlocal potentials of Ref.~\cite{Entem:2017gor} at the same chiral order, and
  their much smaller residual cutoff dependence, see Fig.~17 of
  Ref.~\cite{Reinert:2017usi}, can presumably be traced back to the improved
  semilocal regulator, which maintains the long-range part of the
  interaction as described in section~\ref{sec41}. We also addressed
  in detail the issue of uncertainty quantification in the NN sector. 
In particular, we discussed statistical uncertainties of NN and
$\pi$N LECs and their propagation to selected observables as well as
uncertainty introduced by fixing the maximum fit energy in the
determination of the NN LECs. We also estimated truncation errors at
various chiral orders using the Bayesian model specified in section~\ref{sec5}. 
\item
Beyond the NN sector, calculations based on the SMS interactions have so far been
carried out up to N$^2$LO \cite{Epelbaum:2019zqc}. The LECs $c_D$ and $c_E$, which enter the
3NF at this order, have been determined from the $^3$H binding energy
and the very precise pd cross section data at $E_{\rm lab}^N = 70$~MeV
from Ref.~\cite{Sekiguchi:2002sf}. Using the employed Bayesian model to estimate truncation
uncertainties, the predicted ground state energies of p-shell nuclei
up to $A = 16$ are generally in a good agreement with the
data.  Also the predicted Nd scattering observables including the
vector analyzing power $A_y$ are consistent with
experimental data within errors. We performed an additional test of the
employed Bayesian model for truncation errors by exploring the impact of
selected short-range 3NF terms at N$^4$LO on observables in Nd elastic
scattering. Our results suggest that a high-precision description of
Nd scattering data will likely require the chiral expansion of the 3NF
to be pushed to N$^4$LO.  
\item
The novel semilocal nuclear forces, derived in the finite-cutoff
formulation of chiral EFT with short-range interactions counted
according to NDA (i.e.~the Weinberg scheme), have already
been successfully confronted with few-nucleon data and 
passed a number of a-posteriori consistency checks as briefly
summarized below:
\begin{itemize}

\item[--] Using the minimal basis of the order-$Q^4$ NN contact
   interactions as detailed in section~\ref{sec6}, the LECs determined from the np and pp scattering data
   come out of a natural size, see Fig.~\ref{fig:NaturalLECs}. The same is true for
   the LECs $c_D$ and $c_E$ entering the leading 3NF, as can be seen
   e.g.~from the corresponding expectation values in the $^3$H state
   \cite{Epelbaum:2019zqc}.
 \item[--] The residual cutoff-dependence of NN phase shifts is
   strongly reduced at N$^{3,4}$LO as compared to N$^{1,2}$LO within the
   considered $\Lambda$-range, see e.g.~Fig.~4 of
   Ref.~\cite{Epelbaum:2015pfa}.
 \item[--] There is a clear systematic improvement in the
   description of np and pp data with increasing chiral
   orders, see Tab.~\ref{tab:chi2_AllOrders}. At order $Q^3$ (i.e.~N$^2$LO), 
   this improvement results solely from taking into account the
   parameter-free subleading TPEP contributions.
   Notice that certain alternative power counting
   schemes suggest that
   some of the contact interactions that appear at order $Q^4$ in the
   Weinberg scheme are enhanced and should yield  contributions
   to observables larger than the
   order-$Q^3$ TPEP, see e.g.~Table~1 of
   Ref.~\cite{Griesshammer:2015osb}. The clear evidence of the
   chiral TPEP 
   at orders $Q^3$ and $Q^5$ observed in
   Refs.~\cite{Epelbaum:2014efa,Epelbaum:2014sza,Reinert:2017usi}  does,
   however, not support such 
   alternative scenarios.
 \item[--] The resulting convergence pattern of
   the EFT expansion for selected NN observables was scrutinized using
   Bayesian statistical methods, see section~\ref{sec5} for details. For not too soft cutoffs,
   the assumed breakdown scale of the EFT expansion of $\Lambda_{\rm
     b} \sim 600$~MeV \cite{Epelbaum:2014efa} was found 
   to be statistically consistent \cite{Furnstahl:2015rha}, see also
   Refs.~\cite{Melendez:2017phj,Epelbaum:2019zqc} for a related discussion.
 \item[--] Scheme-dependence 
   of nuclear potentials offers yet another way to perform nontrivial consistency
   checks of the theoretical framework by explicitly verifying
   (approximate) scheme-independence of observables. In the
   formulation we employ, scheme dependence of the nuclear forces
   first appears at N$^3$LO and manifests itself in their dependence on
   arbitrary real phases 
   $\bar
   \beta_{8}$,  $\bar \beta_{9}$, which parametrize the unitary
   ambiguity of the leading relativistic corrections \cite{Kolling:2011mt,Epelbaum:2014efa}, and the
   appearance of three off-shell short-range operators in the
   $^1$S$_0$ and $^3$S$_1$-$^3$D$_1$ channels proportional to the LECs
   $D_{1S0}^{\rm off}$, $D_{3S1}^{\rm off}$ and $D_{\epsilon 1}^{\rm
     off}$ \cite{Beane:2000fi,Furnstahl:2000we,Reinert:2017usi}.  The SMS potentials
   of Ref.~\cite{Reinert:2017usi} make use of the
   standard choice for $\bar \beta_{8,9}$ namely $\bar
   \beta_8 = - \bar \beta_9 = 1/4$, which
   minimizes the amount of $1/m^2$-corrections to the OPEP,
   and employ $D_{1S0}^{\rm off} = D_{3S1}^{\rm off} = D_{\epsilon 1}^{\rm
     off} = 0$.  Different choices of these parameters lead to
   different off-shell behaviors of the potential. They are related
   to each other by unitary transformations which, however, also
   induce an infinite tower of higher-order terms beyond the order one
   is working. The residual dependence of observables
   on $\bar \beta_{8,9}$ and $D_{i}^{\rm off}$, therefore, probes the
   impact of neglected higher-order terms and should lie within the estimated
   truncation errors.  We have redone the fits at N$^4$LO$^+$ for
   $\Lambda = 450$~MeV using alternative choices of $D_{i}^{\rm
     off}$ \cite{Reinert:2017usi} and also developed a version of the
   potential with $\bar  \beta_8 = \bar \beta_9 = 1/2$
   \cite{Filin:2019eoe}. The letter choice is motivated by the vanishing isoscalar
   exchange charge density operator at N$^3$LO. In all 
   considered cases, we found negligibly small changes in observables in spite
   of strong changes at the interaction level. 
\item[--] Calculations of three- and more-nucleon
  observables based on solely NN interactions are
  incomplete beyond second order. They do, however, provide information about the
  magnitude of the missing 3NF contributions by assessing the spread
  in results at different orders $Q^{\geq 3}$ and via a comparison of such
  incomplete predictions with experimental data. In
  Ref.~\cite{Binder:2015mbz}, such an analysis was
  performed for Nd scattering observables and selected properties of light
  nuclei using the SCS NN interactions of
  Refs.~\cite{Epelbaum:2014efa,Epelbaum:2014sza}.
  The sizes of the 3NF contributions required to bring such incomplete
  results in agreement with experimental data were found to agree 
  well with expectations based on Weinberg's power counting.
Furthermore, recent calculations by the LENPIC Collaboration which 
include the leading 3NF \cite{Epelbaum:2018ogq,Epelbaum:2019zqc}  show that the
resulting N$^2$LO predictions for observables that have not been used in the determination of the LECs
$c_D$, $c_E$ are generally in a good agreement with the data, see 
section~\ref{sec6}. No indications are found for enhanced
contributions of the 3NF in general and of the $c_D$-term in
particular as suggested in Ref.~\cite{Birse:2009my}.  
\end{itemize}  
\end{itemize}

To summarize, major progress has been achieved in recent
years towards developing chiral EFT into a precision tool for
low-energy nuclear physics. In the NN sector, the latest SMS
interactions at fifth chiral order have already reached the accuracy at or
even below permille level for low-energy observables such as e.g.~the
deuteron asymptotic S-state normalization $A_S$, see section~\ref{NNresults}
for details and further examples. The only essential missing step
in the NN sector 
concerns the inclusion of isospin-breaking interactions up to fifth
chiral order. Work along this line is in progress.  

Pushing the precision frontier
beyond the NN system opens exciting perspectives for low-energy
nuclear theory and will allow one to confront chiral EFT with currently
unsolved problems, such as a quantitative description of 3N
scattering observables \cite{KalantarNayestanaki:2011wz}. This,
however, will require to address  the two core challenges:
\begin{itemize}
\item[(i)]
Derivation of \emph{consistent regularized} three- and four-nucleon
forces and exchange charge and current operators at and beyond N$^3$LO as
detailed in section~\ref{subsec:Reg}. This issue has not been paid
attention to in the recent calculations involving the 3NFs
\cite{Golak:2014ksa,Drischler:2017wtt,Hoppe:2019uyw,Huther:2019ont}
and
exchange electroweak currents
\cite{Piarulli:2012bn,Baroni:2016xll,Baroni:2018fdn}  at N$^3$LO. 
As explained in section~\ref{subsec:Reg}, using ad hoc regularization approaches
at N$^3$LO and beyond generally leads to incorrect results for the
scattering amplitude and other observables due to the appearance of uncontrolled short-range
artifacts, which violate chiral symmetry and are not suppressed by
inverse powers of $\Lambda$. This puts the findings of these studies into question.
\item[(ii)]
Determination of the LECs in the 3NF at N$^4$LO. While the N$^3$LO
contributions to the 3NF and 4NF do not involve unknown parameters,
the N$^4$LO corrections to the 3NF  involve 10 LEC
accompanying purely short-range operators \cite{Girlanda:2011fh} and one or more LECs
entering the one-pion-exchange-contact topology, which has not 
been worked out yet. As discussed in section~\ref{3Nresults}, the
determination of these LECs from 3N data will require a computationally
challenging analysis.  
\end{itemize}

As a first example of a precision calculation not restricted to NN
scattering, we have recently determined the deuteron structure radius
with an accuracy below the permille level, $r_{\rm str} = 1.9731^{+0.0013}_{-0.0018}$~fm, by pushing the chiral
expansion of the electromagnetic exchange charge density beyond
N$^3$LO
and performing a thorough analysis of various types of
uncertainty \cite{Filin:2019eoe}.
By combining the predicted value for $r_{\rm str}$
with the very accurate atomic data from isotope shift measurements,
it was, for the first time, possible to extract the neutron charge radius
from experimental data on light nuclei. This study was facilitated by the
absence of loop contributions in the isoscalar exchange charge
density at N$^3$LO \cite{Kolling:2009iq,Kolling:2011mt}, which allowed for a trivial construction
of the corresponding consistently regularized expressions for the
charge operator.
Rederivation of the contributions to 3NFs, 4NFs and
exchange currents at and beyond N$^3$LO using a regulator, consistent
with the one employed in Ref.~\cite{Reinert:2017usi}, would open the way for 
performing similar precision calculations for a broad class of
low-energy few-nucleon reactions.


\section*{Funding}
This work was supported by BMBF
(Grant No. 05P18PCFP1) and by  DFG through funds provided
to the Sino-German CRC 110 110 ``Symmetries and the Emergence of Structure
in QCD'' (Grant No. TRR110).

\section*{Acknowledgments}
We are grateful to Ashot Gasparyan, Jambul Gegelia and
Ulf-G.~Mei{\ss}ner for a careful reading of the
manuscript and useful  comments
and to all members of the LENPIC Collaboration for sharing
their insights into the topics addressed in this review article. We
also thank Pieter Maris for providing us with a source file of Fig.~8 from
Ref.~\cite{Epelbaum:2018ogq}.

\bibliographystyle{frontiersinHLTH&FPHY} 




\end{document}